\documentclass[twocolumn]{IEEEtran}
\usepackage{cite}
\usepackage{color,xcolor}
\usepackage{amsmath,amssymb,amsfonts}
\usepackage{algorithmic}
\usepackage{graphicx}
\usepackage{float}
\usepackage{epstopdf}
\usepackage{textcomp}
\usepackage{multirow}
\usepackage{booktabs}
\usepackage{subfigure}
\usepackage{stfloats}
\usepackage{color}
\usepackage{bm}
\usepackage{makecell}
\def\BibTeX{{\rm B\kern-.05em{\sc i\kern-.025em b}\kern-.08em
    T\kern-.1667em\lower.7ex\hbox{E}\kern-.125emX}}

\begin{document}
\title{{ An SIE Formulation with Triangular Discretization and Loop Analysis for Parameter Extraction of Arbitrarily Shaped Interconnects}}
\author{\IEEEauthorblockN{Zekun Zhu, \IEEEmembership{Graduate Student Member, IEEE}, Zhizhang Chen, \IEEEmembership{Fellow, IEEE}, \\and  Shunchuan Yang, \IEEEmembership{Senior Member, IEEE}\\}

\thanks{Manuscript received xxx; revised xxx.}
\thanks{This work was supported in part by the National Natural Science Foundation of China under Grant 62141405 and Grant 62071125, in part by the Defense Industrial Technology Development Program under Grant JCKY2019601C005, in part by the Pre-Research Project under Grant J2019-VIII-0009-0170, in part by the Fundamental Research Funds for the Central Universities under Grant YWF-22-L-842, and in part by the Fundamental Research Project under Grant 514010505-305. {\it {(Corresponding author: Shunchuan Yang)}}
	
 Zekun Zhu and Shunchuan Yang are with the Research Institute for Frontier Science and the School of Electronic and Information Engineering, Beihang University, Beijing, 100083, China. (e-mail: scyang@buaa.edu.cn and zekunzhu@buaa.edu.cn)
	
Zhizhang Chen is currently with the College of Physics and Information Engineering, Fuzhou University, Fuzhou, Fujian 350108, China, on leave from the Department of Electrical and Computer Engineering, Dalhousie University, Halifax, Nova Scotia, Canada B3H 4R2  (email: zz.chen@ieee.org).
}

}

\maketitle

\begin{abstract}
 A surface integral equation (SIE) formulation under the magneto-quasi-static assumption is proposed to efficiently and accurately model arbitrarily shaped interconnects in packages. Through decently transferring all electromagnetic quantities into circuit elements, the loop analysis is used to carefully construct matrix equations with an independent and complete set of unknowns based on graph theory. In addition, an efficient preconditioner is developed, and the proposed formulation is accelerated by the pre-corrected Fast Fourier Transform (pFFT). Four practical examples, including a rectangular metallic interconnect, bounding wire arrays, interconnects in a real-life circuit and the power distribution network (PDN) used in packages, are carried out to validate its accuracy, efficiency and scalability. Results show that the proposed formulation is accurate, efficient and flexible to model complex interconnects in packages.
\end{abstract}


\section{Introduction}
Interconnects are widely used in 2.5D/3D packages and integrated circuits (ICs) [\citen{PackagingICs_1}][\citen{PackagingICs_2}]. The power integrity/signal integrity (SI/PI) are vital to be considered in design of interconnects in high frequency region for its performance [\citen{PackagingICs_1}]-[\citen{PackagingICs_4}]. Circuit parameters, like parasitic resistance and inductance, are essential to the design of SI/PI [\citen{PackagingICs_2}][\citen{SIPI_1}].

To accurately and efficiently extract those parameters of complex interconnects in packages, many efforts have been made in the last few decades. Several volumetric integral equation (VIE) formulations are proposed to extract the wideband resistance and inductance. FastHenry is one popular open-source solver based on the magneto-quasi-static (MQS) VIE formulation in conjugate with a mesh analysis, and currents are assumed to mainly flow in the longitude direction [\citen{FastHenry}]. However, since electric currents significantly crowd towards surfaces of conductors in high frequency region, extremely fine meshes are usually required to accurately model the skin effect in highly lossy interconnects. Then, several full-wave VIE formulation are proposed to model interconnects under exterior electromagnetic waves or complex inhomogeneous media [\citen{FullWaveVIE_1}]-[\citen{FullWaveVIE_3}].

To improve the computational efficiency, surface integral equation (SIE) formulations are proposed to extract parameters of interconnects. Those formulations usually show performance improvements over their volumetric counterparts, since unknowns only reside on surfaces of interconnects. For example, several full-wave methods based on the SIE formulations are developed in [\citen{SIEFullwave1}]-[\citen{SIEFullwave6}], in which the tangential electric and magnetic fields are calculated on the surfaces of objects. Many formulations based on the SIEs and circuit theory, namely the partial element equivalent circuit (PEEC) methods, are proposed to model circuit structures [\citen{SIEPEEC_0}]-[\citen{SIEPEEC_8}]. In those formulations, electromagnetic quantities are interpreted as circuit elements, and then external excitations are applied for impedance extraction. In addition, an open-source solver, namely FastImp [\citen{FastImp}], based on the mixed potential integral equation (MPIE) formulation was developed to model arbitrarily shaped conductors [\citen{MPIE1}]-[\citen{MPIE3}]. In [\citen{MQSSIE}], an MQS SIE formulation is proposed to model interconnects with rectangular panels. Compared with other formulations, it is more efficient since the surface impedance boundary condition used to model the skin effect. However, currents are still assumed to flow in the longitude direction. Arbitrary currents cannot be easy to be modeled in complex interconnects. The Ansys Q3D extractor based on the MQS SIE formulation was developed to model interconnects in industrial applications [\citen{AnsysQ3D}].

In this article, an SIE formulation under the MQS assumption is proposed to efficiently and accurately model arbitrarily shaped interconnects in packages. To conveniently apply the charge conservation condition and external voltage sources, all electromagnetic quantities are decently transferred into circuit elements. For practical complex interconnects, the equivalent circuits may be nonplanar circuits. Therefore, a loop analysis rather than the traditional mesh analysis is developed to carefully construct matrix equations with an independent and complete set of unknowns based on graph theory. In addition, the pre-corrected Fast Fourier Transform (pFFT) [\citen{PFFT2}][\citen{PFFT}] is successfully implemented to solve matrix equation for large-scale interconnects, and an efficient preconditioner is developed to accelerate the convergence. We carried out four practical examples from simple to complex structures, including a rectangular metallic interconnect, bounding wire arrays, interconnects in a real-life circuit and the power distribution networks (PDNs) used in packages, to validate its accuracy, efficiency and scalability.

Compared with other existing techniques, contributions in this article are mainly in three aspects.
\begin{enumerate}
	\item  An SIE formulation under the MQS assumption with the triangular discretization is used to model arbitrarily shaped interconnects in packages. Through carefully interpreting the SIE formulation as an equivalent circuit, the loop analysis is successfully developed to apply the charge conservation condition and exterior excitations.
	
	\item An efficient preconditioner is introduced to accelerate the convergence of the proposed SIE formulation, and the pFFT algorithm is specially tailored to fast calculate the matrix-vector product for large-scale problems, Therefore, it can model practical complex interconnects. 
	
	\item Four practical complex structures in packages are used to verify the accuracy, efficiency and flexibility of the proposed formulation. Results show that it can efficiently model large-scale interconnects, and solve the practical problems in the real-life circuits with the same level of accuracy compared with the industrial solver.
\end{enumerate}

The article is organized as follows. In Section II, the proposed SIE formulation with the triangular discretization and the modified Rao-Wilton-Glisson (RWG) basis functions are detailed presented. Then, its equivalent circuit corresponding to the SIE formulation is introduced and detailed explained. In Section III, the loop analysis with graph theory is used to apply the charge conservation condition and voltage sources. An effective preconditioner is proposed to accelerate the convergence and the pFFT algorithm is detailed shown in Section IV. Then, four numerical examples are carried out to validate the accuracy and efficiency of the proposed formulation in Section V. Finally, we draw some conclusions in Section VI.

\section{The SIE Formulation for Lossy Interconnects}
\subsection{The MQS SIE formulation}
Without loss of generality, external excitations are not included inside interconnects. The electric field in the exterior space can be expressed as
\begin{equation}\label{MPIE1}
	\mathbf{E} = -j\omega\mathbf{A}-\nabla\Phi,
\end{equation}
where $\mathbf{E}$, $\mathbf{A}$, $\Phi$ are the electric field, the vector potential and the scalar potential, respectively. $\omega$ denotes the angular frequency. $\mathbf{A}$ can be expressed in terms of the electric current density and the Green's function as
\begin{equation}\label{VectorPotential}
	\mathbf{A} = \mu_{0}\int G_{0}\left(\mathbf{r},\mathbf{r}'\right)\mathbf{J}\left(\mathbf{r}'\right)d\mathbf{r}'.
\end{equation}
By substituting (\ref{VectorPotential}) into (\ref{MPIE1}), we have
\begin{equation}\label{MPIE2}
	\mathbf{E} = -j\omega\mu_{0}\int G_{0}\left(\mathbf{r},\mathbf{r}'\right)\mathbf{J}\left(\mathbf{r}'\right)d\mathbf{r}'-\nabla\Phi,
\end{equation}
which is the well-known MPIE. In this article, we consider that the skin effect is well-developed in the high frequency region. Therefore, electric currents mainly flow towards the perimeter of interconnects. The surface impedance operator can be used to relate the tangential electric field to the surface current density, which is given by 
\begin{equation}\label{ImpedanceOp}
	Z_{s} = \sqrt{j\omega\mu/\sigma},
\end{equation}
where $\sigma$ is the conductivity of the interconnect. It should be noted that (\ref{ImpedanceOp}) is only valid in the high frequency region, where the skin effect is well-developed. When parameters in the low frequency region are considered, other impedance operators, such as the generalized impedance boundary condition [\citen{GIBC_1}]-[\citen{GIBC_3}], should be used. 

By considering lossy effects imposed by (\ref{ImpedanceOp}), (\ref{MPIE2}) can be modified as
\begin{equation}\label{MPIE3}
	Z_{s}\mathbf{J}\left(\mathbf{r}\right)+j\omega\mu_{0}\int G_{0}\left(\mathbf{r},\mathbf{r}'\right)\mathbf{J}\left(\mathbf{r}'\right)d\mathbf{r}' = -\nabla\Phi.
\end{equation}
(\ref{MPIE3}) is the SIE formulation to model lossy interconnects.

Since interconnects in packages are considered in this article, of which the large typical sizes are only several millimeters, the MQS assumption can be safely used. Therefore, the static Green's function in free space is used in (\ref{MPIE3}), which is given by
\begin{equation}\label{StaticGF}
	G_{0}\left(\mathbf{r},\mathbf{r}'\right) = \frac{1}{4\pi\lvert\mathbf{r}-\mathbf{r}'\rvert}.
\end{equation}
The MQS assumption also implies the electric charge conservation condition, namely $\nabla\cdot\mathbf{J} = 0$. In Section III, the loop analysis will be selected to enforce it.

\subsection{Surface Discretization Through Triangles}
Generally, volumetric or surface elements, such as tetrahedron, filament, triangle and panel, are used to divide interconnects into small ones to calculate numerical results. Volumetric elements are usually applied in VIEs, such as filaments in FastHenry [\citen{FastHenry}], voxels in VoxHenry [\citen{VoxHenry}], and tetrahedrons in [\citen{FullWaveVIE_1}], which lead to prohibitively large modeling errors of complex structures or computational cost in terms of runtime and memory to model the well-developed skin effect. For surface elements, rectangular panels with pulse basis functions are developed in [\citen{MQSSIE}], in which currents are assumed to flow in the longitude direction. Currents in arbitrary directions are not easy to be supported.

In our implementation, triangles are used to discretize the surface of interconnects. Therefore, arbitrarily shaped interconnects can be easily discretized with considerably small modeling errors. In addition, the RWG functions can be used to support currents in arbitrary direction. Therefore, triangles are much more preferred for complex interconnects.


\subsection{Construction of Matrix Equations Through Modified RWG Basis Functions}

The method of moments (MoM) is chosen to solve the electric current density in (\ref{MPIE3}). Once triangle meshes on the surface of interconnects are constructed, the modified RWG functions are used to discretize surface current density $\mathbf{J}$, which can be expressed as 
\begin{equation}\label{RWG}
	{{\bf{f}}_n}({\bf{r}}) = \left\{ {\begin{array}{*{2}{cc}}
			\frac{1}{{2A_n^ + }}{\bm{\rho }}_n^ + ({\bf{r}}) &{\bf{r}}\,{\rm{ in }}\,T_n^ + \\[0.7em]
			\frac{1}{{2A_n^ - }}{\bm{\rho }}_n^ - ({\bf{r}}) &{\bf{r}}\,{\rm{ in }}\,T_n^ - \\[0.7em]
			0 &otherwise
	\end{array}} \right.
\end{equation}
Compared with the traditional RWG basis function in [\citen{RWGFunction}], it should be noted that the edge length does not exist for (\ref{RWG}) in our implementation. Then, the current density $\mathbf J$ can be expanded using the modified RWG functions
\begin{equation}\label{ExpandJ}
	{\bf{J}}({\bf{r}}) = \sum\limits_{i = 1}^{{N_e}} {{I_{{b_i}}}{{\bf{f}}_i}({\bf{r}})},
\end{equation}
where $N_e$ is the total number of edges. The expansion coefficient $I_{b_{i}}$ denotes the current flows across the edge $i$ from the triangle $T_{n}^{+}$ to $T_{n}^{-}$. Since $\mathbf{f}_n$ is scaled by its corresponding edge length, $I_{b_{i}}$ denotes the electric current rather than the current density flowing through the edge $i$ as the original RWG basis function. Such modification is imported and makes the loop analysis be easily applied as shown in Section III.

In addition, the pulse functions are used to expand $\Phi$, which is expressed as
\begin{equation}\label{ExpandPotential}
\Phi ({\bf{r}}) = \sum\limits_{i = 1}^{{N_t}} {{\varphi _i}{\Pi _i}({\bf{r}})},
\end{equation}
where $N_t$ is the overall number of triangles, $\Pi _i({\bf{r}})$ is $1$ on the $i$th triangle and 0 outside it, respectively.

We substitute (\ref{ExpandJ}) and (\ref{ExpandPotential}) into (\ref{MPIE3}), and the Galerkin scheme is selected to test (\ref{MPIE3}). The following matrix equation can be obtained 
\begin{equation}\label{BranchVandI}
{{\bf{Z}}_b}{{\bf{I}}_b} = {{\bf{V}}_b},
\end{equation}
where 
\begin{align}
	{\left[ {{{\bf{Z}}_b}} \right]_{ij}} &=  \frac{{j\omega \mu }}{{4\pi }}\int\limits_{{S_i}} {{{\bf{f}}_i}({\bf{r}})} \int\limits_{{S_j}} {\frac{{{{\bf{f}}_j}({\bf{r'}})}}{{\left| {{\bf{r}} - {\bf{r'}}} \right|}}d{\bf{r}}d{\bf{r'}}}\notag\\
\label{Zb}	& \quad\quad\quad\quad\quad\quad\quad\quad+{Z_s}\int\limits_{{S_i}} {{{\bf{f}}_i}({\bf{r}}){{\bf{f}}_j}({\bf{r}}')d{\bf{r}}} ,\\
\label{Vb}	{\left[ {{{\bf{V}}_b}} \right]_i}\, &= \varphi _i^ +  - \varphi _i^ - .
\end{align}
$\mathbf{V}_b$ is the potential difference between two triangles shared with the edge $i$, and $\mathbf{I}_b$ collects all unknowns in a column vector. To accurately solve (\ref{BranchVandI}), the modified nodal analysis (MNA) [\citen{NodaA}] or the mesh analysis [\citen{FastHenry}] is required to enforce the charge conversation condition. This part will be introduced in Section III.

\subsection{Physical Interpretation of (\ref{BranchVandI}) in the View of Circuit Theory}
In fact, under the MQS assumption, (\ref{BranchVandI}) can be completely interpreted through circuit theory, which makes it easy and convenient to enforce the charge conservation condition and external excitations.
\begin{figure}
	\centering
	\includegraphics[width=0.5\textwidth]{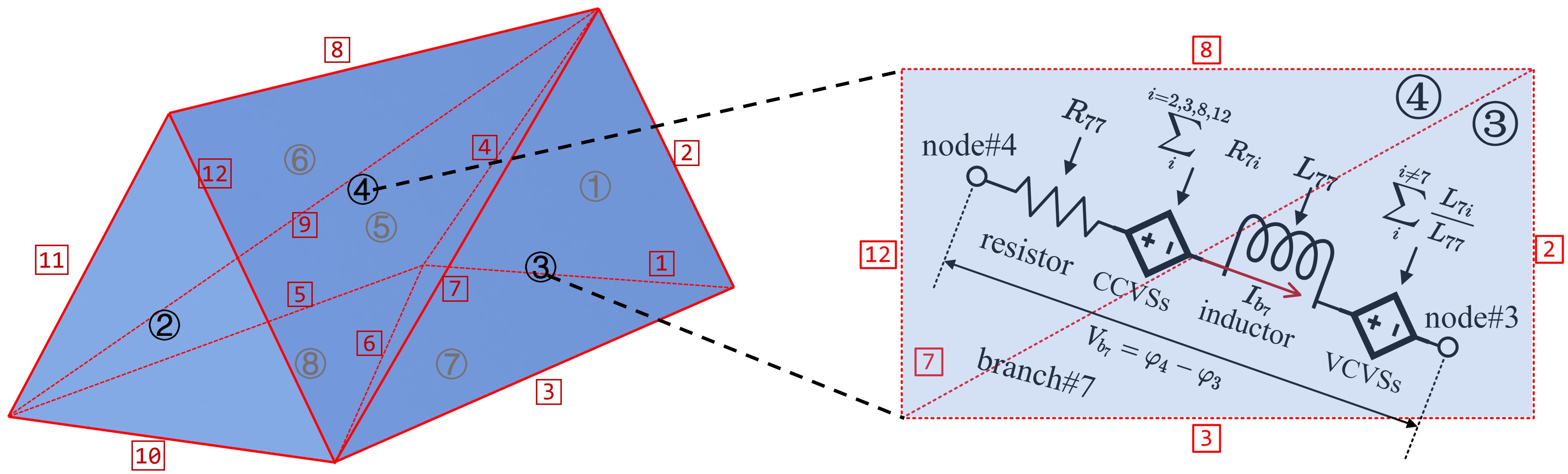}
	\caption{The triangle discretization for a triangular prism interconnect, and the equivalent circuit for a triangle pair.}
	\label{CDinTp}
\end{figure}
To make the derivation clear, we introduce two domains: the electromagnetic domain and the circuit domain. As shown in Fig. \ref{CDinTp}, each triangle is treated as one circuit node in the circuit domain, and one edge between two triangles represents a circuit branch between two nodes. Under these intuitive interpretations, $\mathbf{I}_b$ and $\mathbf{V}_b$ can be regarded as branch currents and voltages in the circuit domain.

In (\ref{Zb}), $\mathbf{Z}_b$ can be separated into two parts: the resistance $\mathbf{R}_b$ and the inductance $\mathbf{L}_b$. $\mathbf{R}_b$ represents the first term of $\mathbf{Z}_b$, and $\mathbf{L}_b$ is the second term. The equivalent circuits for Row\#$i$ of $\mathbf{Z}_b$ are in Fig. \ref{CDinTp}. In the circuit domain, $\left[\mathbf{L}_b\right]_{ii}$ represents the self-inductance, while $\left[\mathbf{L}_b\right]_{ij}\left(j=1,2,\ldots,n,i\neq j\right)$ denotes voltage controlled voltage sources (VCVSs). Similarly, $\left[\mathbf{R}_b\right]_{ii}$ can be represented by a resistor $R_{ii}$, and other four terms of $\left[\mathbf{R}_b\right]_{ij}$ are treated as four current controlled voltage sources (CCVSs) [\citen{CircuitEq}], which are serially connected with $R_{ii}$. Table  \ref{EMDomain2CCDomain} summarizes the relationship between quantities in the electromagnetic domain and their corresponding physical concepts or elements in the circuit domain.

\begin{figure}
	\centering
	\includegraphics[width=0.48\textwidth]{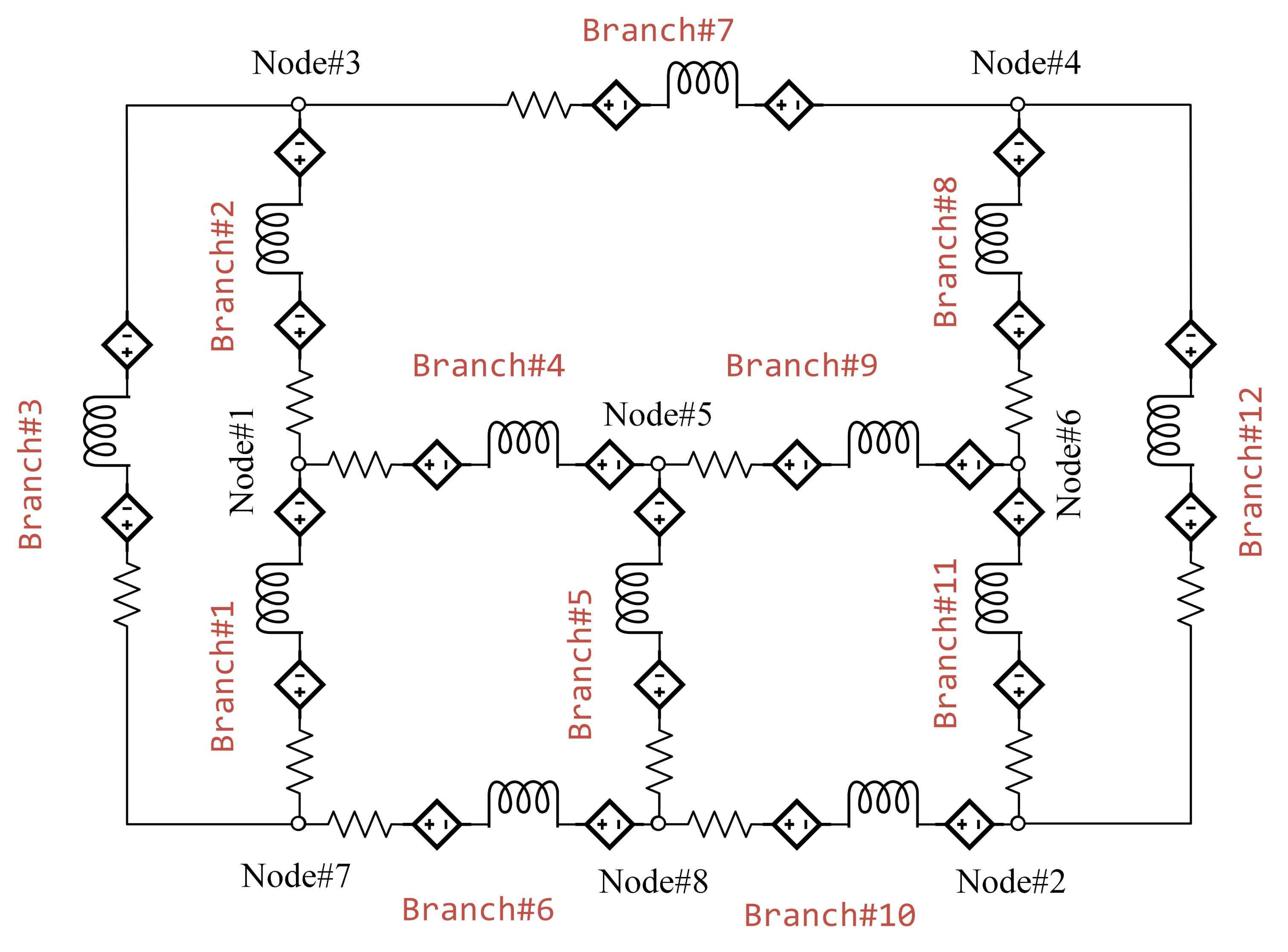}
	\caption{The circuit corresponds to the interconnect in Fig. \ref{CDinTp} by applying the relationship between electromagnetic quantities and circuit elements in Table  \ref{EMDomain2CCDomain}.}
	\label{EqC}
\end{figure}

Fig. \ref{CDinTp} gives a simple example, which is a prism-shaped interconnect. 8 triangular patches are used to construct its surface, which are marked through black numbers with circles. There are 12 edges in total and the red numbers with squares are used to mark them. The equivalence in Table \ref{EMDomain2CCDomain} is applied to this interconnect, and a planar circuit is obtained as shown in Fig. \ref{EqC}. Therefore, to this point, it is fully interpreted as a circuit. In Section III, we will use graph theory to select an independent unknown set, and then solve it.

\renewcommand\arraystretch{1.2}
\begin{table}
	\centering
	\caption{Relationship between quantities in  the electromagnetic domain and elements in the circuit domain}\label{EMDomain2CCDomain}
	\begin{tabular}{|l|c|}
		\hline
		\textbf{EM domain}                              &\textbf{Circuit domain} \\
		\hline 
		Triangle                                                  &Node   \\
		\hline
		Edge                                                       &Branch            \\
		\hline
		$\left[\mathbf{I}_{b}\right]$                    &Branch current\\
		\hline
		$\left[\mathbf{V}_{b}\right]$                    &Branch voltage\\
		\hline
		$\left[\mathbf{L}_{b}\right]_{ii}$               &Indcutor \\
		\hline
		$\left[\mathbf{L}_{b}\right]_{ij}\left(i\neq j\right)$               &VCVS\\
		\hline
		$\left[\mathbf{R}_{b}\right]_{ii}$               &Resistor \\
		\hline
	    $\left[\mathbf{R}_{b}\right]_{ij}\left(i\neq j\right)$              &CCVS \\
		\hline
	\end{tabular}
\end{table}

\section{Loop Analysis}
In Section II, (\ref{BranchVandI}) is derived to describe the relationship between the branch current and voltage. To solve it, additional constraints are required in the practical applications, such as the charge conservation law and external excitations. The MNA [\citen{NodaA}], which enforces the Kirchoff's current law, was widely used in many applications [\citen{CircuitEq}]. For each node (triangle), summation of currents flowing into and out of one circuit node should be zero, which leads to an additional matrix equation upon branch currents. Therefore, unknowns in the MNA include branch currents and node voltages. The dimension of the final matrix equation is the summation of the number of edges and triangles. Another option is to use the mesh analysis, which can enforce the Kirchoff's voltage law, and leads to a much smaller matrix equation compared with that from the MNA. However, the mesh analysis is only applicable for planar circuit, which are not true for practical complex interconnects. To overcome this issue, we propose to use the loop analysis to enforce those additional conditions for general applications.

\begin{figure}
	\centering
	\includegraphics[width=0.4\textwidth]{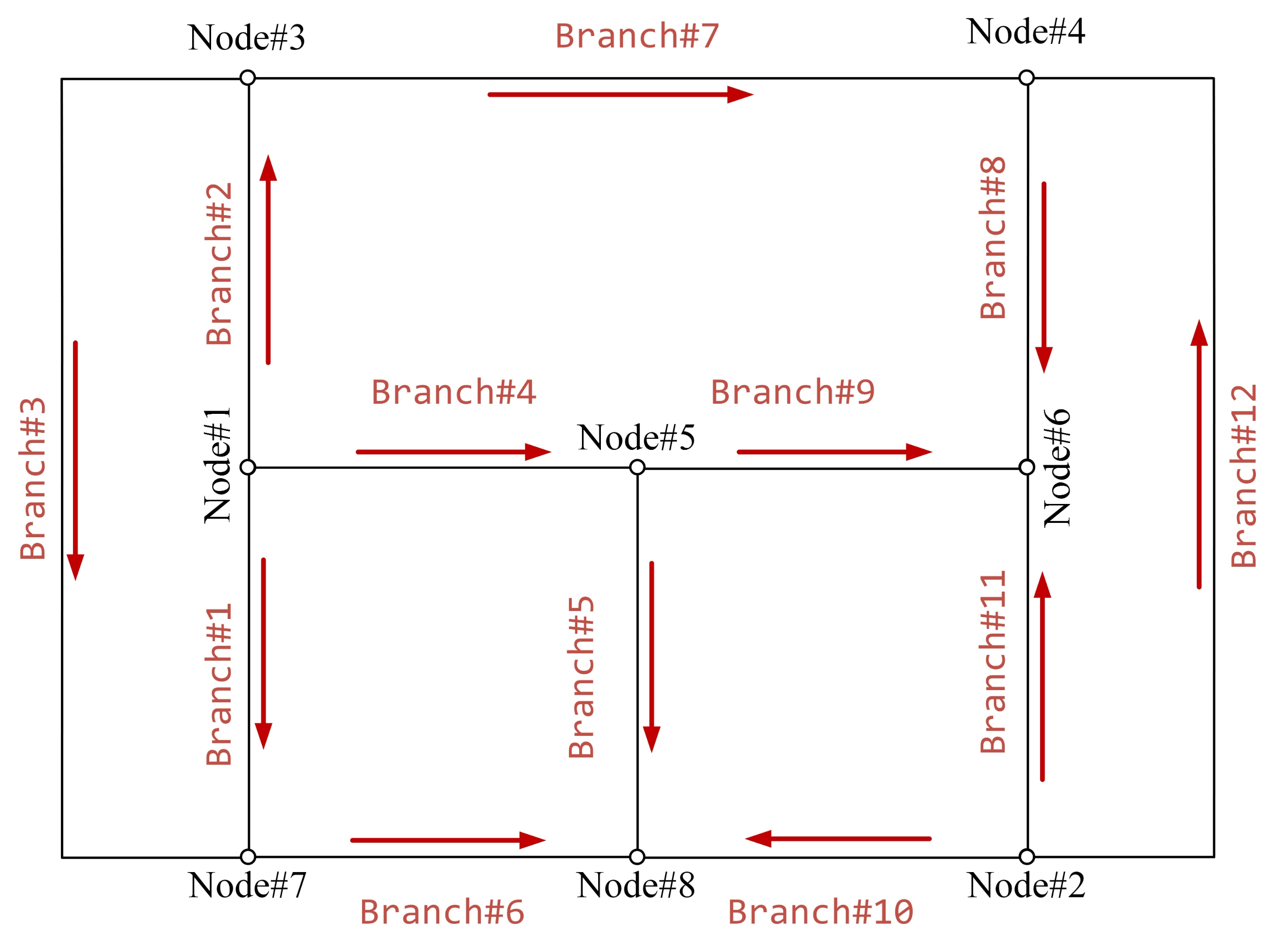}
	\caption{The graph for the circuit in Fig. \ref{EqC} and the direction of branch currents.}
	\label{TopoGraph}
\end{figure}

\begin{figure}
	\centering
	\includegraphics[width=0.499\textwidth]{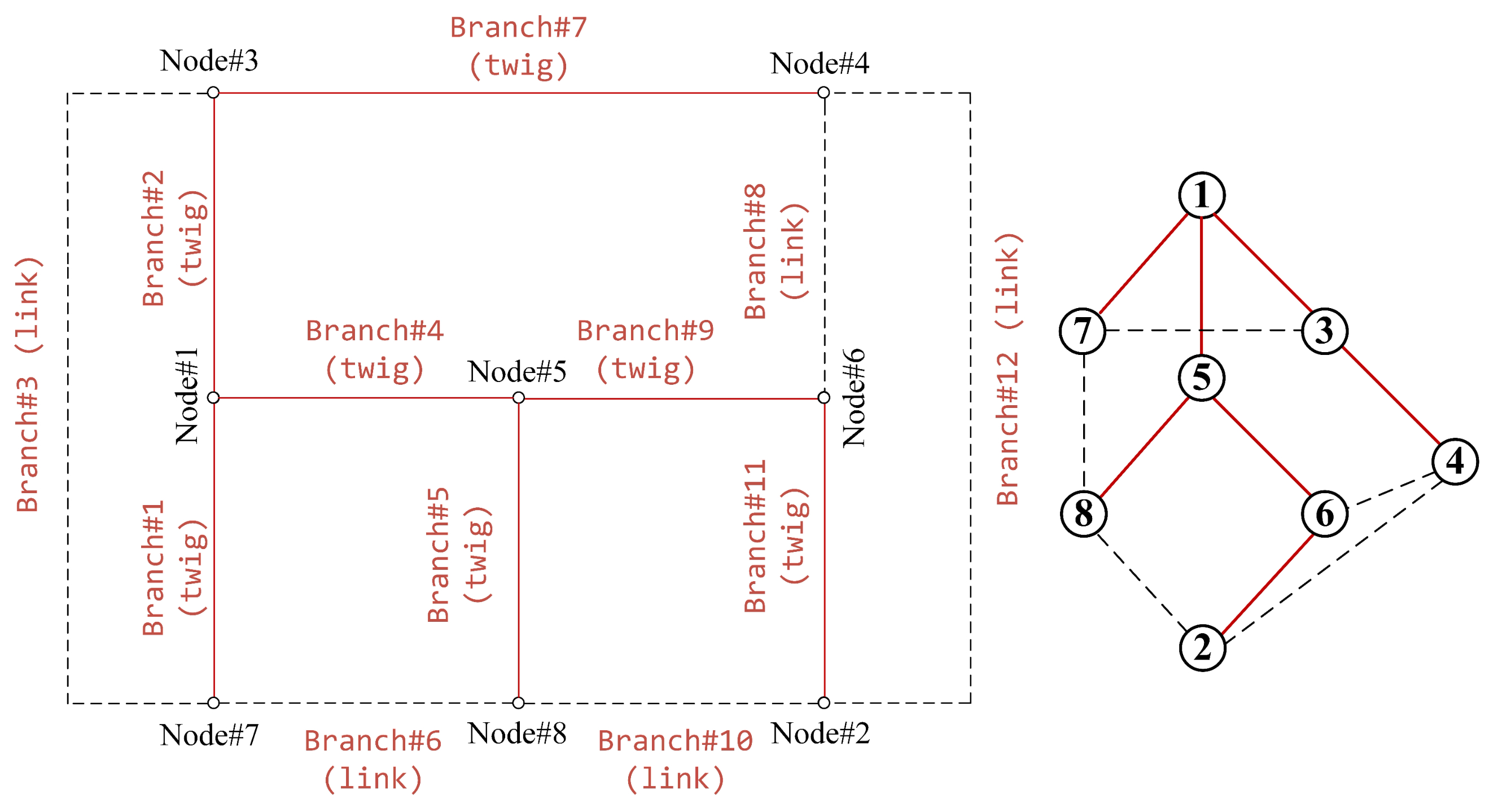}
	\caption{A tree for the graph in Fig. \ref{TopoGraph} and the corresponding twigs and links. The right panel shows the tree connection.}
	\label{TreeDefinition} 
\end{figure}


\subsection{Graph Theory in the Circuit Analysis}
This subsection briefly introduces some essential concepts used in our analysis, and then uses them to solve (\ref{BranchVandI}). When all elements of the equivalent circuit in Fig. \ref{EqC} are removed, a graph of its topological connection can be obtained as shown in Fig. \ref{TopoGraph}. Our following analysis is based on this graph. According to graph theory [\citen{GraphT1}], each connected graph can construct a {\it tree}, which is defined by all nodes connecting through branches, but without including any closed loops. Once a tree for the circuit is generated, all branches in the graph are divided into two groups: those in or not in the tree. Branches in the tree are called {\it twigs}, and the other branches are called {\it links}. It should be noted that the tree for a graph may have many possibilities. Fig. \ref{TreeDefinition} shows one possible tree for the graph in Fig. \ref{TopoGraph}, in which twigs are in red solid lines and links are in block dash lines.

Obviously, if there are $n$ nodes and $b$ branches for a graph with one degree of separation, the overall count of twigs should be $(n-1)$ and the overall count of links is $(b-n+1)$. The {\it degree of separation} is the number of completely separated parts, which denotes the number of fully separated conductors without any physically connection. Therefore, the overall counts of twigs and links are $(n-s)$ and $(b-n+s)$ for a graph with $s$ degree of separation, respectively. 

\begin{figure}
	\centering
	\includegraphics[width=0.4\textwidth]{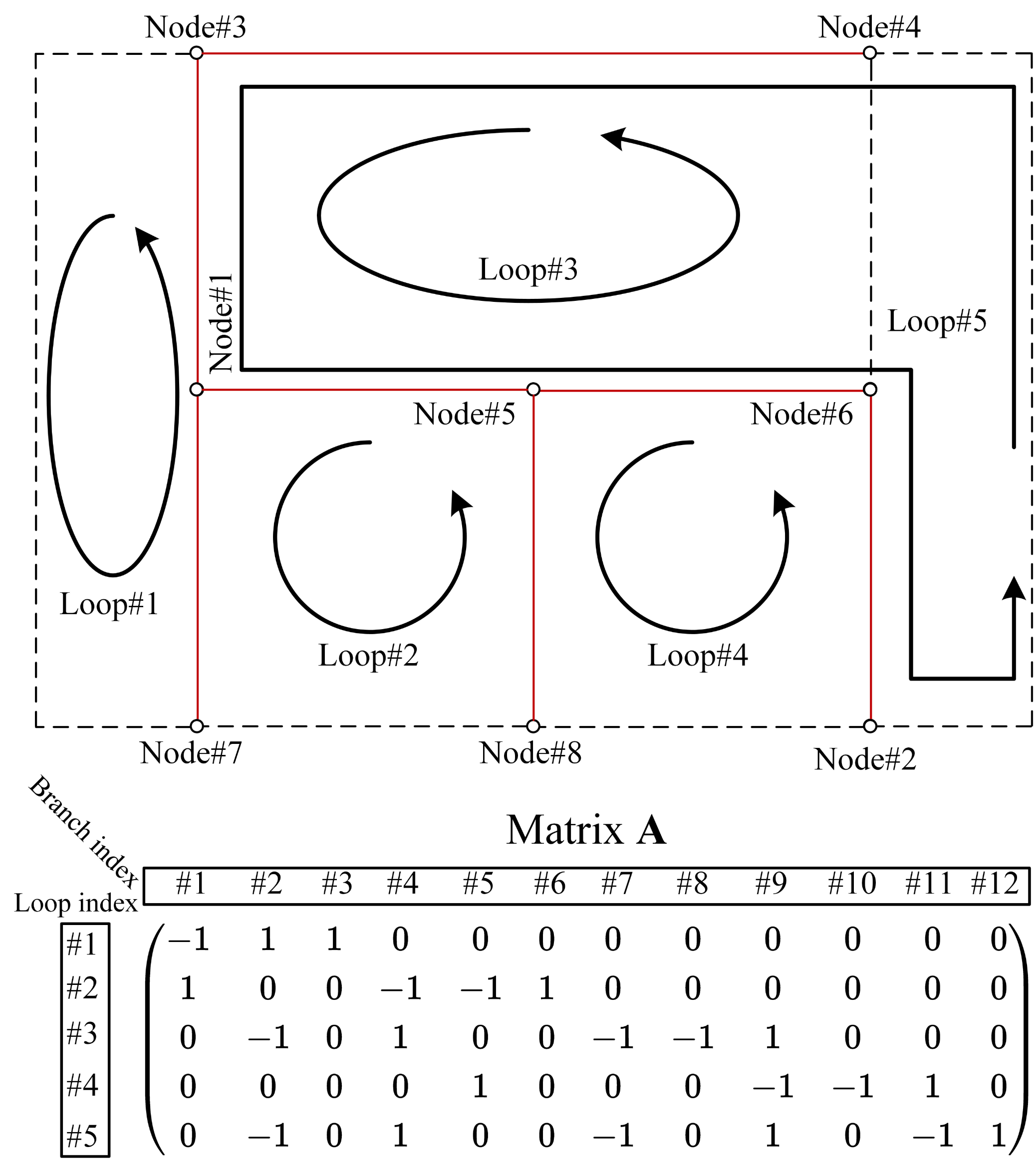}
	\caption{The fundamental loops for the selected tree without additional source, and the corresponding transfer matrix $\mathbf A$.}
	\label{LoopIllu} 
\end{figure}

In the loop analysis, loop currents rather than branch currents are required to be solved. The loops cannot be arbitrarily chosen, and the set of loops should be independent and complete. Generally, one possible choice is to generate the {\it fundamental loop}, which is a loop consisting of only one link and several connected twigs (see Fig. \ref{LoopIllu}). Therefore, the overall count of fundamental loops is equal to that of links. It is found that if loop currents are defined on the fundamental loops, they are independent with each other, and a set including all fundamental loops is complete. Therefore, we can easily construct such set through a tree for the graph. In Fig. \ref{LoopIllu}, all the black solid circles consist of such set.

\subsection{Transfer Branch Quantities to Loop Counterparts}
To get loop currents and voltages, the Kirchoff’s voltage law is applied in each fundamental loop. For each loop, the summation of branch voltages must be equal to the loop voltage, which is expressed as
\begin{equation}\label{VolT}
	\mathbf{A} \mathbf{V}_b=\mathbf{V}_l,
\end{equation}
where $\mathbf{A}$ transfers the branch voltage $\mathbf{V}_b$ to the loop counterpart $\mathbf{V}_l$. Generally, the loop voltage is zero if there are no additional voltage sources in the corresponding loop. The dimension of $\mathbf{A}$ is $l \times b$, where $b$ is the overall count of branches and $l = (b-n+s)$ is the number of links. Its entries are $-1$ or $1$, which corresponds to the direction of loops and branch voltage drops. Fig. \ref{LoopIllu} gives elements of $\mathbf{A}$ for the corresponding loops. $\mathbf{A}$ is a sparse matrix and can be efficiently stored through the compressed column storage (CCS) or compressed row storage (CRS) format.

The transfer matrix from loop currents to branch counterparts is the transpose of $\mathbf{A}$, which is given by
\begin{equation}\label{CurT}
	\mathbf{A}^T \mathbf{I}_l=\mathbf{I}_b,
\end{equation}
where $\mathbf{I}_l$ is a column vector including all loop currents. By substituting (\ref{VolT}) and (\ref{CurT}) into (\ref{BranchVandI}), we have
\begin{equation}\label{LoopVandI}
	\mathbf{Z}_l \mathbf{I}_l=\mathbf{V}_l,
\end{equation}
where $\mathbf{Z}_l=\mathbf{A Z}_b \mathbf{A}^T$.

By using $\mathbf{A}$ and $\mathbf{A}^T$, branch quantities are replaced by their corresponding loop counterparts, and the dimension of the matrix equation significantly reduces from $(b+n)$ to $(b-n+s)$. For a closed surface, the number of edges is always one and a half times as the number of triangles. Therefore, the dimension of matrix equation generated by the loop analysis is about 20\% of that by the MNA.

\begin{figure}
	\centering
	\includegraphics[width=0.4\textwidth]{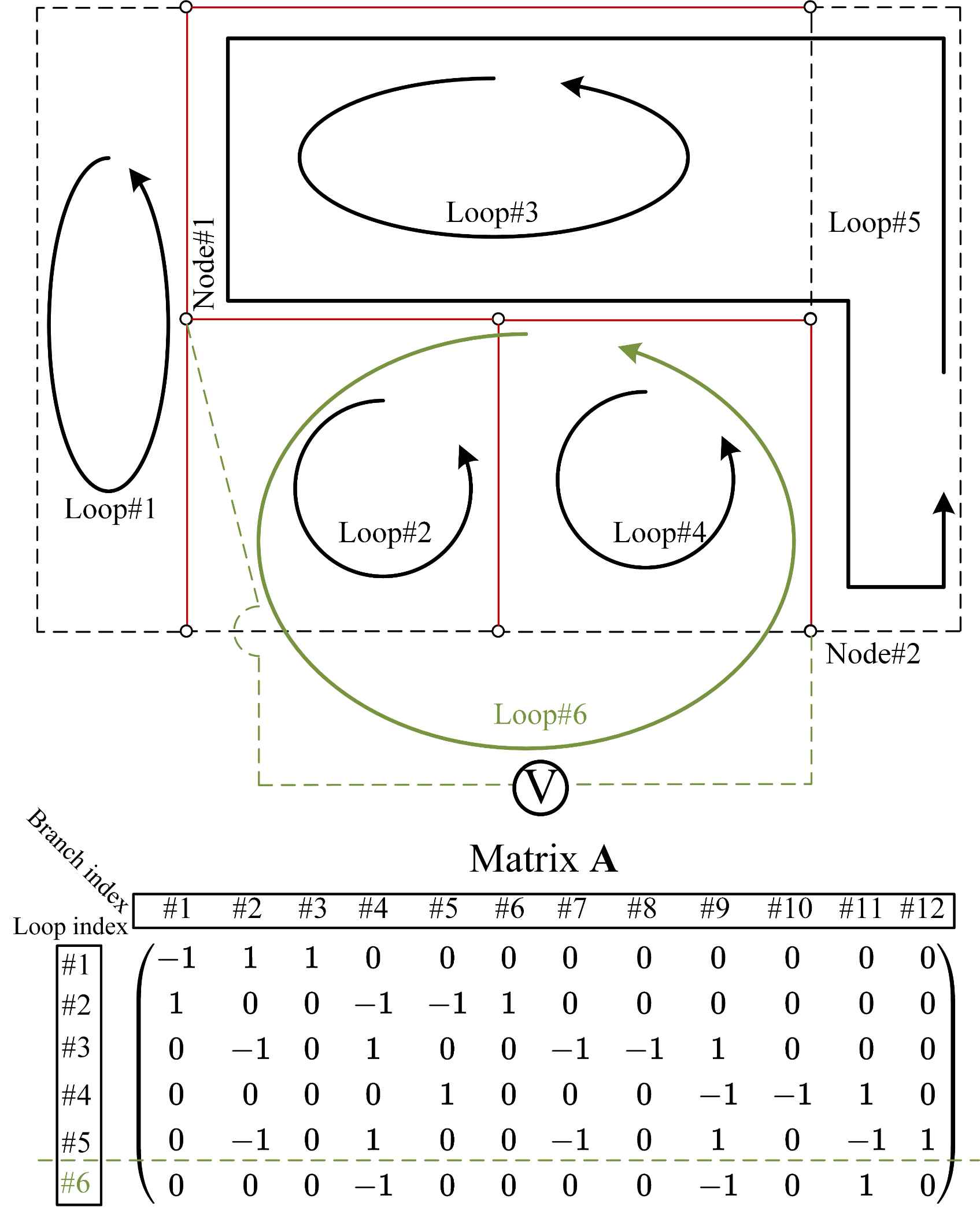}
	\caption{The loops with a voltage source between Node\#1 and Node\#2 and the modified Matrix $\mathbf{A}$.}
	\label{LoopWithEx} 
\end{figure}
\subsection{Apply Voltage Sources to the Circuit}

Generally, a voltage source is attached between two circuit nodes. The tree of a circuit does not need to be modified. It is also true for twigs, because no additional circuit nodes are added. The only difference is that an additional link is generated. Therefore, a new fundamental loop should be considered as shown in Fig. \ref{LoopWithEx}. The dimension of matrix in (\ref{LoopVandI}) is $(b-n+s+p)$ if external voltage sources are considered, where $p$ is the overall number of exterior voltage sources. The modified matrix $\mathbf A$ is shown in Fig. \ref{LoopWithEx}.

Once the loop analysis is carried out, and the external voltage source is attached to the circuit, the final matrix equation is expressed as 
\begin{equation}\label{FinalME}
	{\bf{A}}{{\bf{Z}}_b}{{\bf{A}}^T}{{\bf{I}}_l} = {{\bf{V}}_l}.
\end{equation}

\section{Preconditioning and PFFT Acceleration}
When large-scale interconnects are considered, (\ref{FinalME}) have to be solved with iterative algorithms. The generalized minimal residual method (GMRES) [\citen{GMRES}] is used to solve (\ref{FinalME}) in our implementation. To speed up its convergence, a preconditioner is carefully designed and the pFFT is implemented to accelerate the matrix-vector multiplication. This section introduces the preconditioning matrix and the pFFT acceleration algorithm.

\subsection{Preconditioning}
To reduce the overall iteration number and accelerate the convergence, an efficient preconditioner is required. In this article, the preconditioner is defined as
\begin{equation}\label{Precond}
	{\bf{P}} = {\mathbf{AZ}}_b^N{{\bf{A}}^T},
\end{equation}
where $\mathbf{Z}_b^N$ is a diagonal matrix, in which the entries are obtained from $\mathbf{Z}_b$. Therefore, $\mathbf{P}$ is a symmetric matrix. The left preconditioning technique is applied to (\ref{LoopVandI}), and (\ref{LoopWithP}) can be obtained as
\begin{equation}\label{LoopWithP}
	{{\bf{P}}^{ - 1}}{{\bf{Z}}_l}{{\bf{I}}_l} = {{\bf{P}}^{ - 1}}{{\bf{V}}_l}.
\end{equation}
It is obvious that the inverse of $\mathbf P$ should be effectively calculated when the GMRES is used to solve (\ref{FinalME}). Fortunately, since $\mathbf{A}$ and $\mathbf{A}^T$ are sparse matrices and $\mathbf{Z}_b^N$ is a diagonal matrix, $\mathbf{P}$ is a sparse matrix. A direct factorization algorithm can be used to efficiently calculate the inverse of $\mathbf{P}$, such as the PARDISO solver in MKL [\citen{MKL}].

Three matrix-vector products are required to be calculated for (\ref{LoopWithP}). It should be noted that ${\widetilde {\bf{I}}_l} = {{\bf{Z}}_l}{{\bf{I}}_l}$ is very time-consuming, since $\mathbf{Z}_l$ is a full matrix. However, it is easy to calculate ${{\bf{P}}^{ - 1}}{{\bf{V}}_l}$ and ${{\bf{P}}^{ - 1}}{\widetilde {\bf{I}}_l}$ because $\mathbf{P}^{-1}$ can be effectively calculated.   $\mathbf{Z}_b$ can be separated into two parts mentioned in Section II-B. It can be expressed as 
\begin{equation}\label{SepZb}
	{{\bf{Z}}_b} = {\bf{R}}_b + {\bf{L}}_b,
\end{equation}
where ${\bf{R}}_b$ is the first term of (\ref{Zb}), which is a sparse matrix, and ${\bf{L}}_b$ is the second term, which is a full matrix. Therefore, we only need to accelerate ${\bf{AL}}_b{{\bf{A}}^T}{{\bf{I}}_l}$.

\subsection{PFFT Acceleration}
\begin{figure}
	\centering
	\includegraphics[width=0.3\textwidth]{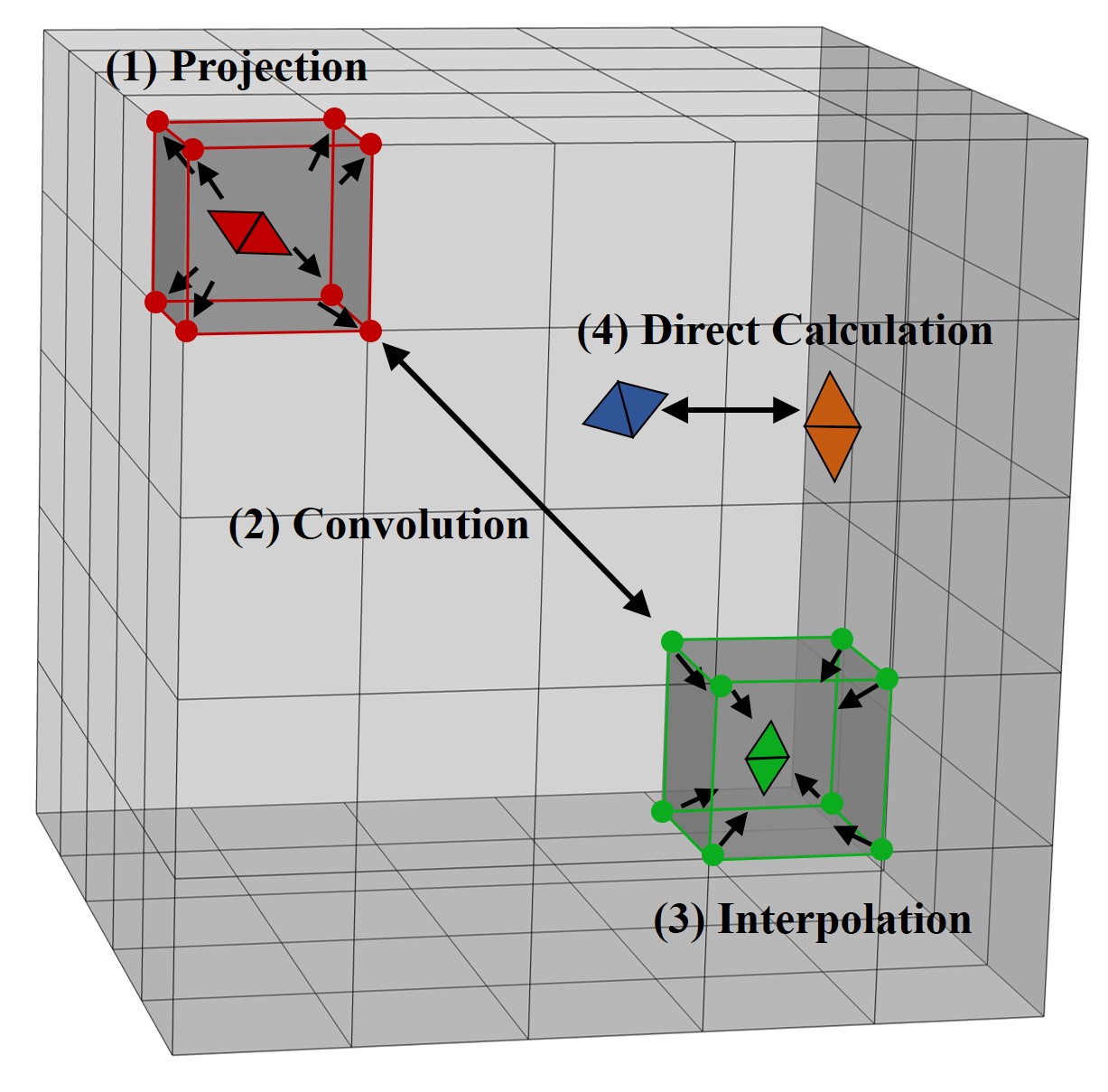}
	\caption{Flowchart through the pFFT algorithm to accelerate the matrix-vector multiplication.}
	\label{pFFTGrid}
\end{figure}

For a Toeplitz matrix, the matrix-vector multiplication can be efficiently computed using the pFFT algorithm, which implies that the time complexity can be reduced from $O\left(N^2\right)$ to $O\left(NlogN\right)$, and the memory cost is reduced to $O\left(N\right)$. By assuming uniformly distributed grids in the three-dimensional space as shown in Fig. \ref{pFFTGrid}, $G_0\left(\mathbf{r},\mathbf{r}'\right)$ is a three-level Toeplitz matrix because the Green’s function is shift-invariant in the three-dimensional space. Therefore, the FFT can be used to efficiently accelerate the matrix-vector multiplication.

Assume  ${\bf{\bar \alpha }} = {{\bf{A}}^T}{{\bf{I}}_l}$, ${\bf{AL}}_b{{\bf{A}}^T}{{\bf{I}}_l}$ can be rewritten as ${\bf{AL}}_b\bf{\bar \alpha} $.
In the pFFT algorithm, the whole computation domain can be categorized into two parts: near- and far-field regions. The pFFT algorithm consists of four steps as follows.
\begin{enumerate}
\item  Construction of the project matrix from a triangle pair to its nearby grids, which can be expressed as
\begin{equation}\label{Projection}
	{\overline {\bf{Q}} _g} = {\bf{B\bar \alpha }},
\end{equation}
where $\mathbf{B}$  is the projection matrix, and ${\overline {\bf{Q}} _g} $  is equivalent point currents on the uniform grids. The collection of these grids is defined as a stencil, and the order of stencil $O_s$ is determined by the number of grids $Ng$. In general, $O_s$ grids in the $x$, $y$ and $z$ direction from its nearest grid to the center of the target triangle are selected as a $O_s$-order stencil.

\item  Construction the convolution matrix to efficiently calculate the potential between gird points through the FFT algorithm, which is expressed as
\begin{equation}\label{FFTCal}
	{\overline {\bf{\varphi }} _g} = {\bf{H}}{\overline {\bf{Q}} _g},
\end{equation}
where $\mathbf{H}$ is a Toeplitz matrix. The FFT can be used to efficiently calculate the matrix-vector multiplication ${\bf{H}}{\overline {\bf{Q}} _g}$ with $O\left(NlogN\right)$, where $N$ is the overall count of grid points.

\item  Construction of the interpolation matrix from the nearby grids to the desired triangle, which is expressed as
\begin{equation}\label{interpolation }
	{{\bf{\Phi }}_g} = {\bf{I}}{\overline {\bf{\varphi }} _g},
\end{equation}
where the interpolation matrix $\mathbf{I}$ is the transposition matrix of $\mathbf{B}$ when the Galerkin scheme is applied.

\item  For near-field regions, entries of  $\mathbf{L}_b$ are directly calculated, and are collected in $\mathbf{L}_d$. Therefore,  $\mathbf{L}_b$ can be approximated as
\begin{equation}\label{Lb}
	{\bf{L}}_b \approx {{\bf{L}}_d} + {\bf{IHB}},
\end{equation}
where $\mathbf{L}_d$, $\mathbf{I}$, $\mathbf{B}$  are all sparse matrices. Finally, the matrix-vector multiplication ${\bf{AL}}_b{{\bf{A}}^T}{{\bf{I}}_l}$ is calculated by ${\bf{A}}\left[{{\bf{L}}_d} + {\bf{IHB}}\right]{{\bf{A}}^T}{{\bf{I}}_l}$.
\end{enumerate}

\section{Numerical Results And Discussion}
In this section, four numerical examples are computed using the proposed formulation and the industrial solver Ansys Q3D, which was developed based on the SIE formulation for complex structures. The resistance and inductance are calculated, and results are compared with those from the Ansys Q3D. In addition, the effectiveness of the pFFT algorithm and the proposed preconditioner are discussed and verified. Our in-house solver is developed with C++ without any parallel computations, and all simulations are carried out on the workstation with a 3.2 GHz CPU and 1 TB memory.

\subsection{One Rectangular Interconnect}
\begin{figure}
	\centering
	\includegraphics[width=0.4\textwidth]{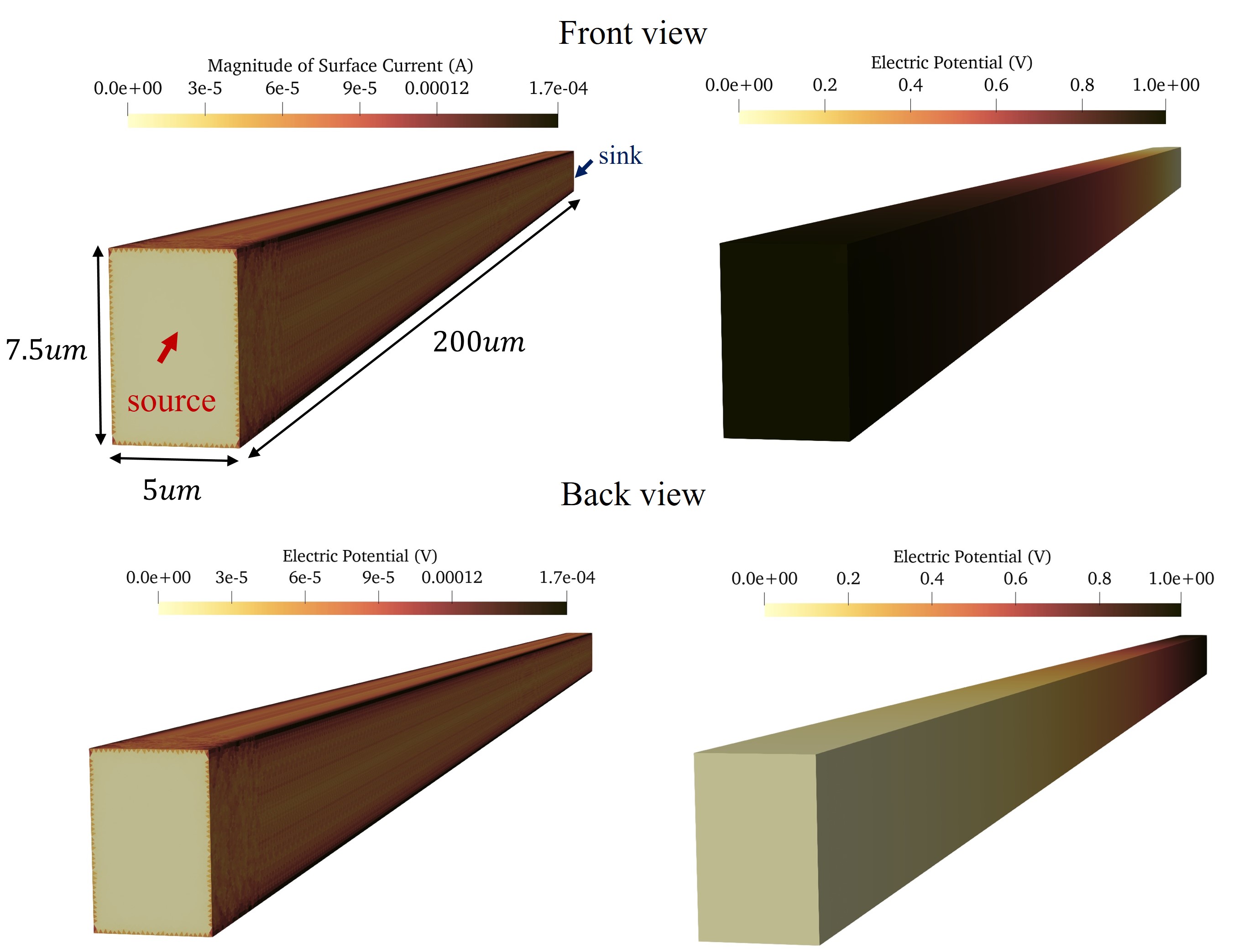}
	\caption{A rectangular copper interconnect and the excitation configuration. The magnitude of surface current and electric potential at 100 GHz are illustrated.}
	\label{Case1-Model}
\end{figure}
\begin{figure}
	\centering
	\includegraphics[width=0.4\textwidth]{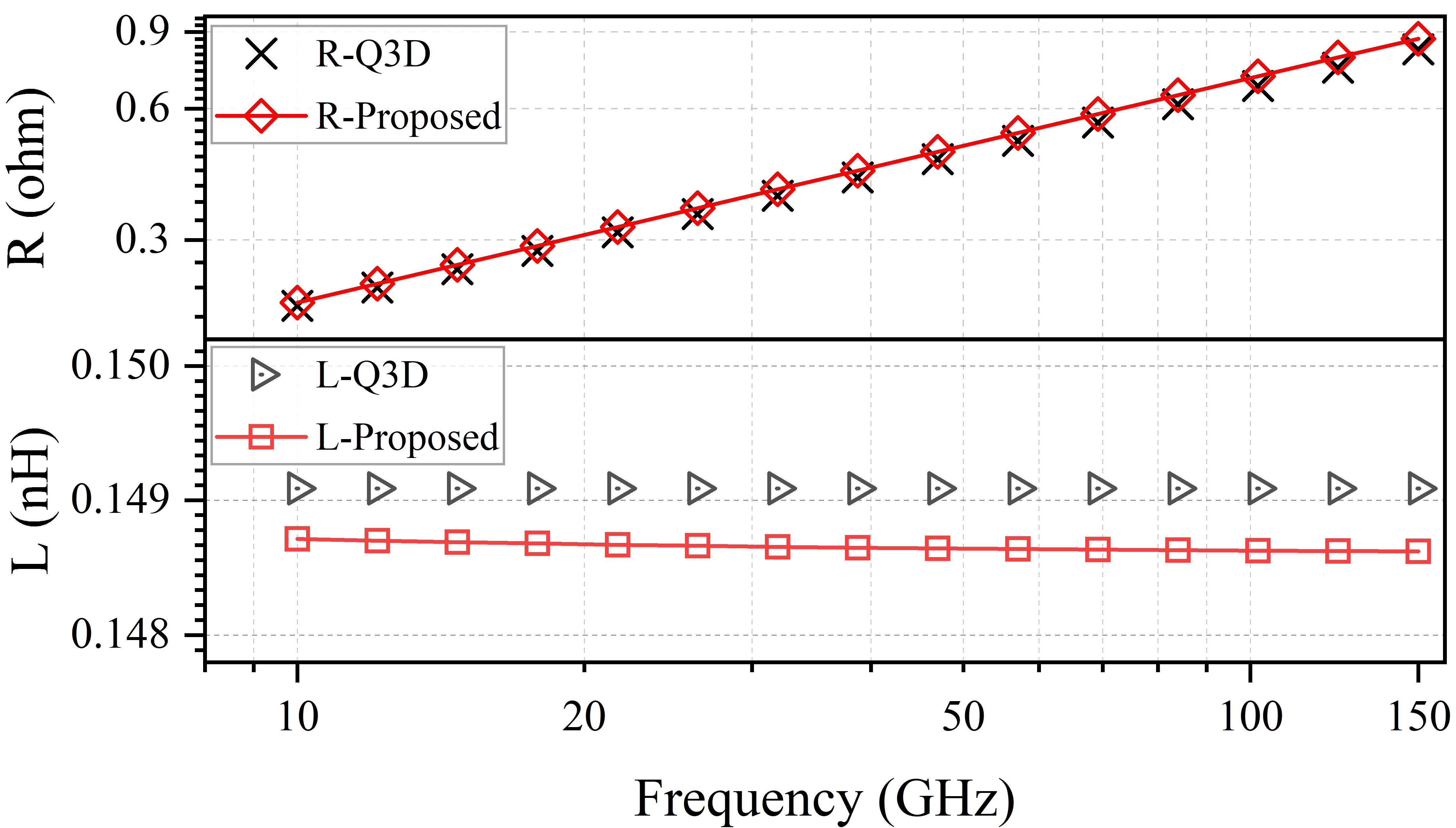}
	\caption{The resistance and inductance obtained from the Ansys Q3D and the proposed formulation.}
	\label{Case1_RL}
\end{figure}
\begin{figure}
	\centering
	\includegraphics[width=0.39\textwidth]{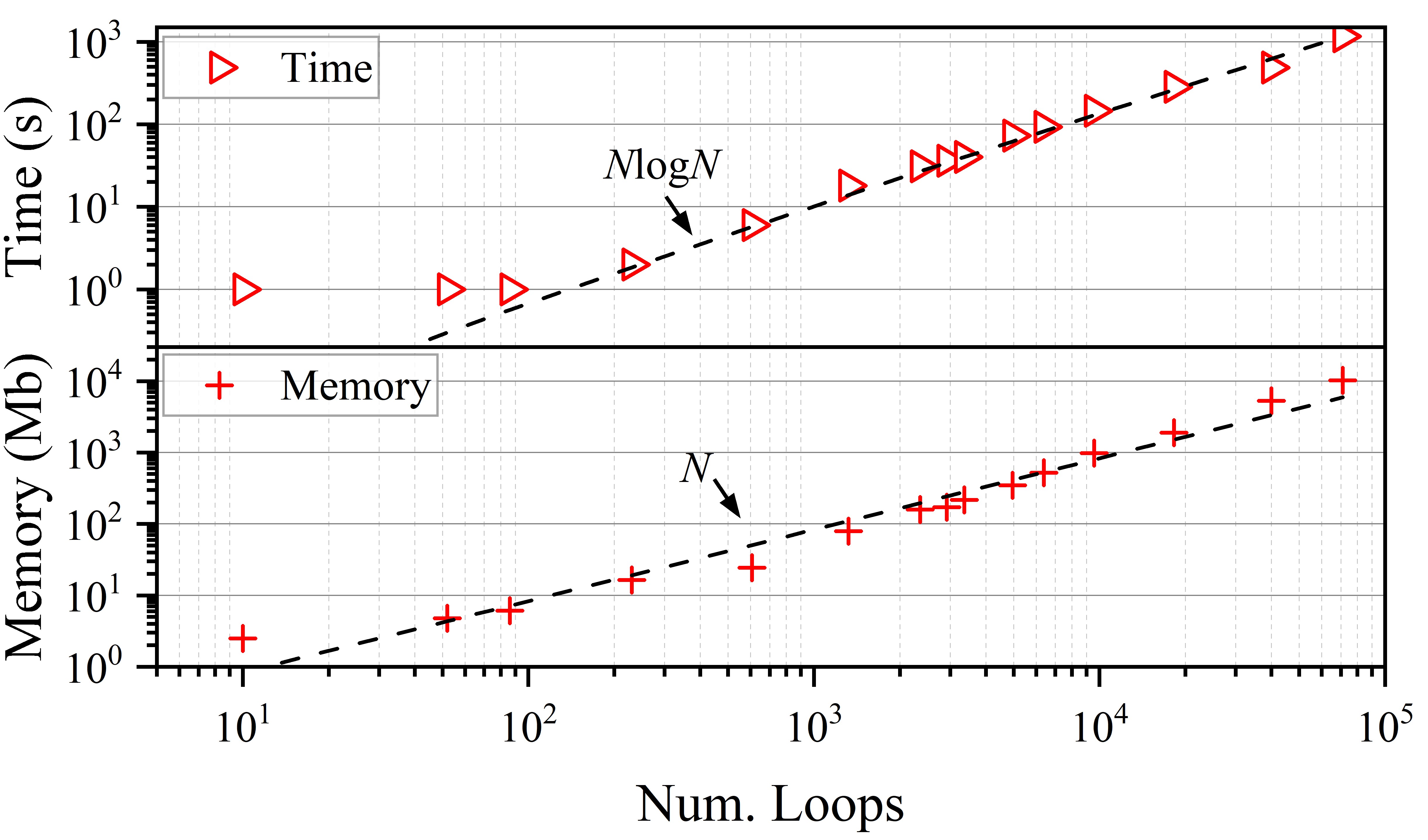}
	\caption{The time and memory consumption when different meshes are used to discretize the interconnect.}
	\label{Case1_TM}
\end{figure}
\begin{figure}
	\centering
	\includegraphics[width=0.4\textwidth]{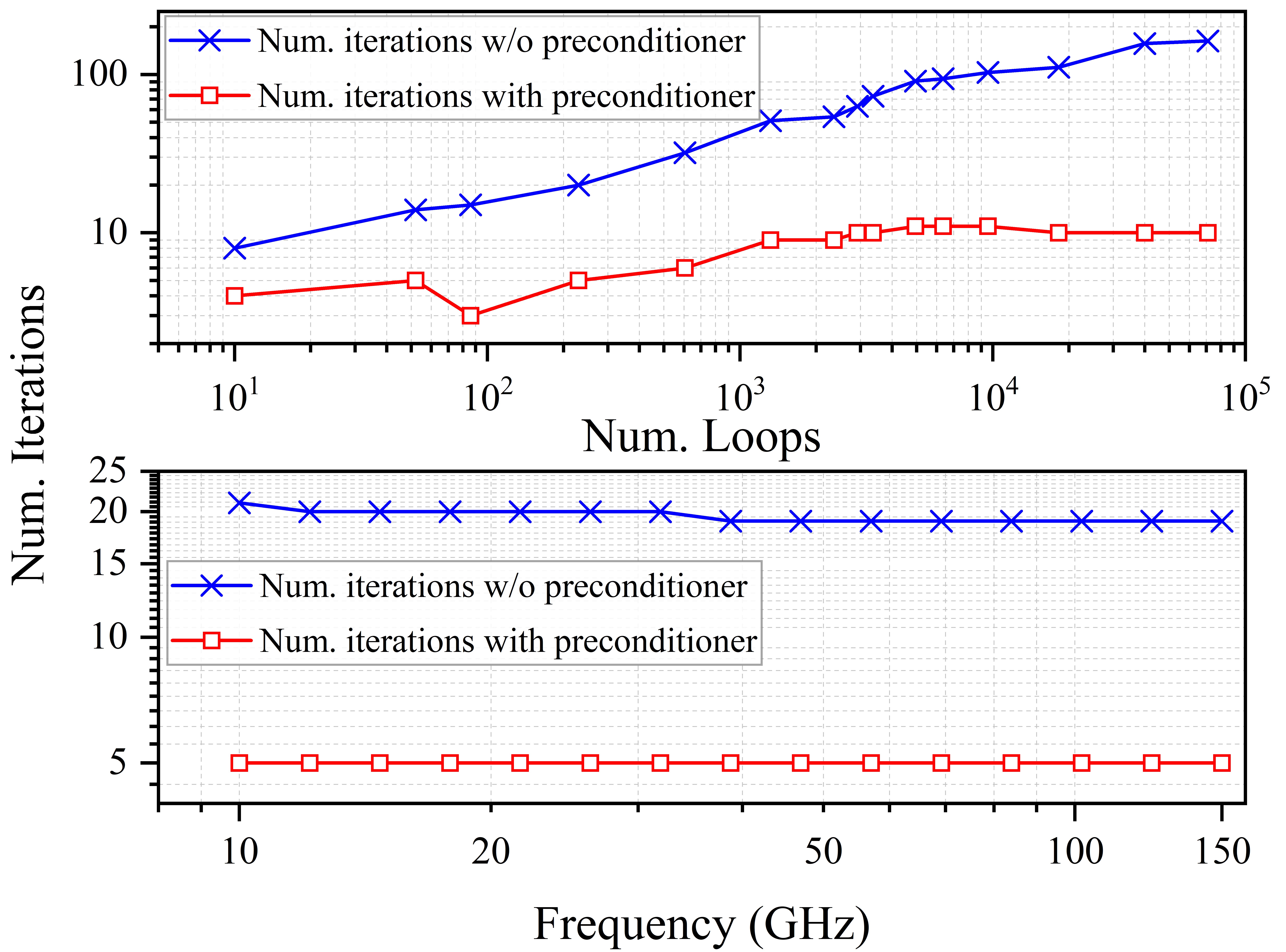}
	\caption{The number of iterations for the GMRES with and without the preconditioner versus the frequency and meshes.}
	\label{Case1_It}
\end{figure}

The first structure is a rectangular copper interconnect with the size of 5 um$\times$7.5 um$\times$200 um as shown in Fig. \ref{Case1-Model}. A voltage source is added between two ends of this interconnect, and the surface with red arrows is the source, which has higher potential compared with that at the far end of the interconnect and indicate currents flow into the interconnect. The corresponding surface with the blue arrow is the sink, and has the lower potential compared with that on the source and represents that currents flow out of the interconnect. In Section II-A, it is mentioned that the surface impedance operation should be used in a limited frequency range, and it works well when the skin depth is less than $1/3$ thickness of interconnects in general. Therefore, its wideband properties between 10 GHz$\sim$150 GHz were studied, where the surface impedance operator can well represent the relationship between the surface current density and the tangential electric field. We used 428 triangles to discretize the surface of this interconnect, and there were 230 loops to construct (\ref{LoopVandI}). Fig. \ref{Case1_RL} shows the resistance and inductance calculated by the Ansys Q3D and the proposed formulation. For the resistance, two results are in good agreement in the whole frequency region. It can be noticed that the inductance does not vary with frequency for the Ansys Q3D. In fact, the formulation used in Ansys Q3D is also the MPIE, and the difference is that the objects are considered as the perfect electric conductors (PECs), which implies that the left hand side of (\ref{MPIE2}) is zero. Therefore, the inductance is independent of frequency for the Ansys Q3D, and the resistance obtained from the proposed formulation shows slightly difference compared with that from the Ansys Q3D at the very high frequency.

The effectiveness of the pFFT algorithm is discussed through the runtime and memory consumption as shown in Fig. \ref{Case1_TM}. We discretized the interconnect using the mesh with different sizes, and recorded the runtime and memory consumption, in which the frequency was set to 100 GHz, which is one of the challenging scenarios in this example. The runtime and memory increase as the number of loops increases. Curves versus the runtime and memory consumption approximately match $O(NlogN)$ and $O(N)$ when the number of unknowns is large enough. When the number of unknowns is very small, some additional operations take up a large portion of the runtime, such as I/O operations.

In addition, the convergence property in the GMRES solver is very important for the efficient parameter extraction. Therefore, we investigated the number of iterations with and without the preconditioner $\mathbf{P}$ in (\ref{Precond}) for the convergent tolerance of $10^{-3}$. As shown in Fig. \ref{Case1_It}, the top subfigure shows the number of iterations when the frequency is set to 100 GHz. As the dimension of the system matrix increases, the number of iterations also increases with or without the preconditioner. However, the preconditioner can approximately reduce the number of iterations by one order when the number of unknowns is large enough. In the bottom subfigure of Fig. \ref{Case1_It}, we used 428 triangles to discretize the interconnect, and the number of iterations with different frequencies were recorded. Similarly, the proposed preconditioner can significantly reduce the overall iteration number, and hence accelerate the convergence.

In Fig. \ref{Case1-Model}, the distribution of electric potential and magnitude of surface current are illustrated. The branch current can be calculated by (\ref{CurT}) since we have obtained the loop current. Similarly, the branch voltages are obtained through (\ref{VolT}). In our simulation, the electric potential on the surfaces of sinks is set as zero, and then the electric potential on each triangular panel can be calculated through the branch voltages.  

\subsection{The Bounding Wire Array}
\begin{figure}
	\centering
	\includegraphics[width=0.4\textwidth]{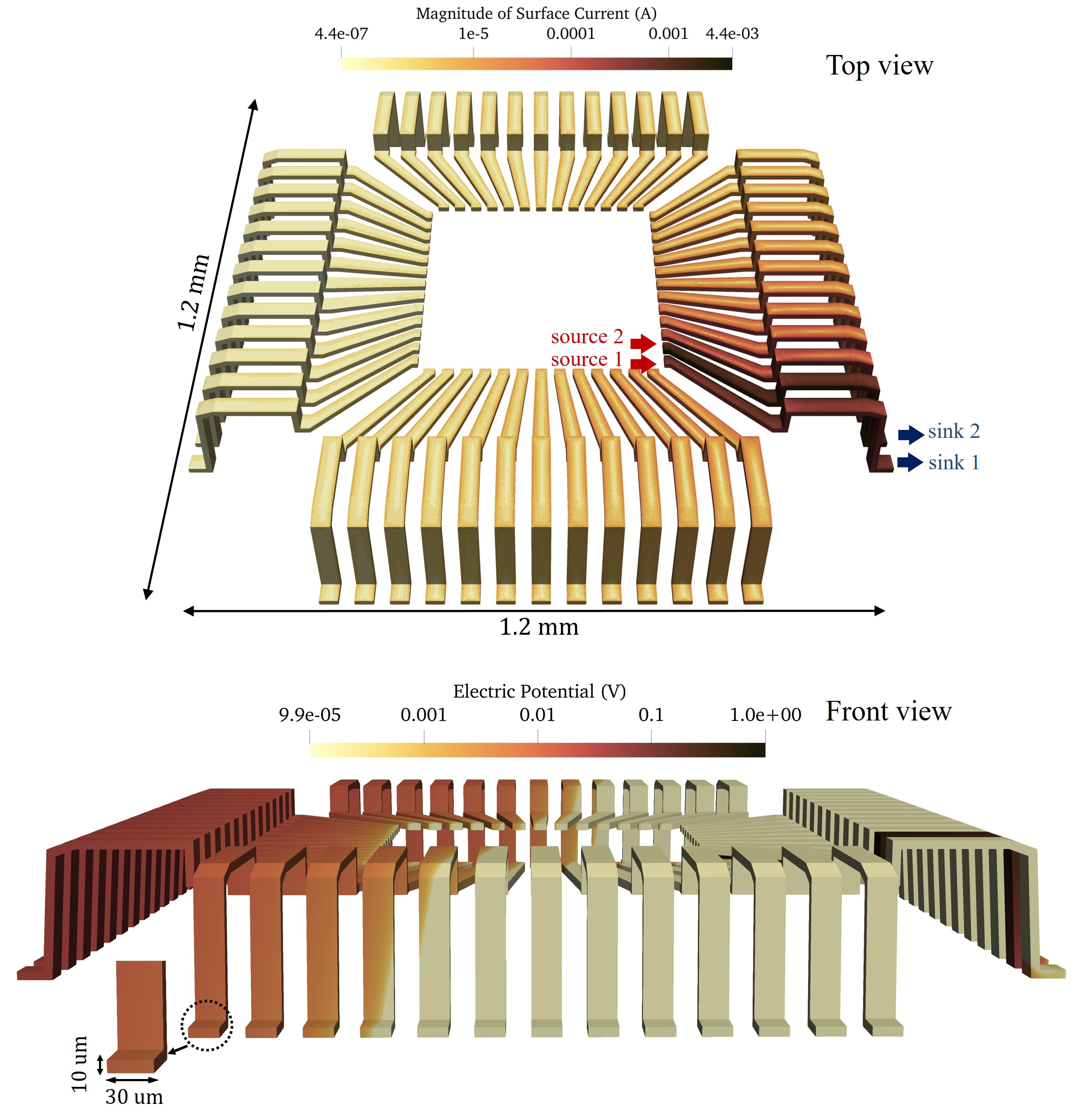}
	\caption{The geometric details and the ports definition used in our simulation. The top subfigure is the magnitude of surface current when Port\#2 is excited, and the bottom subfigure is the electric potential.}
	\label{Case2_Model}
\end{figure}
\begin{figure}
	\centering
	\includegraphics[width=0.4\textwidth]{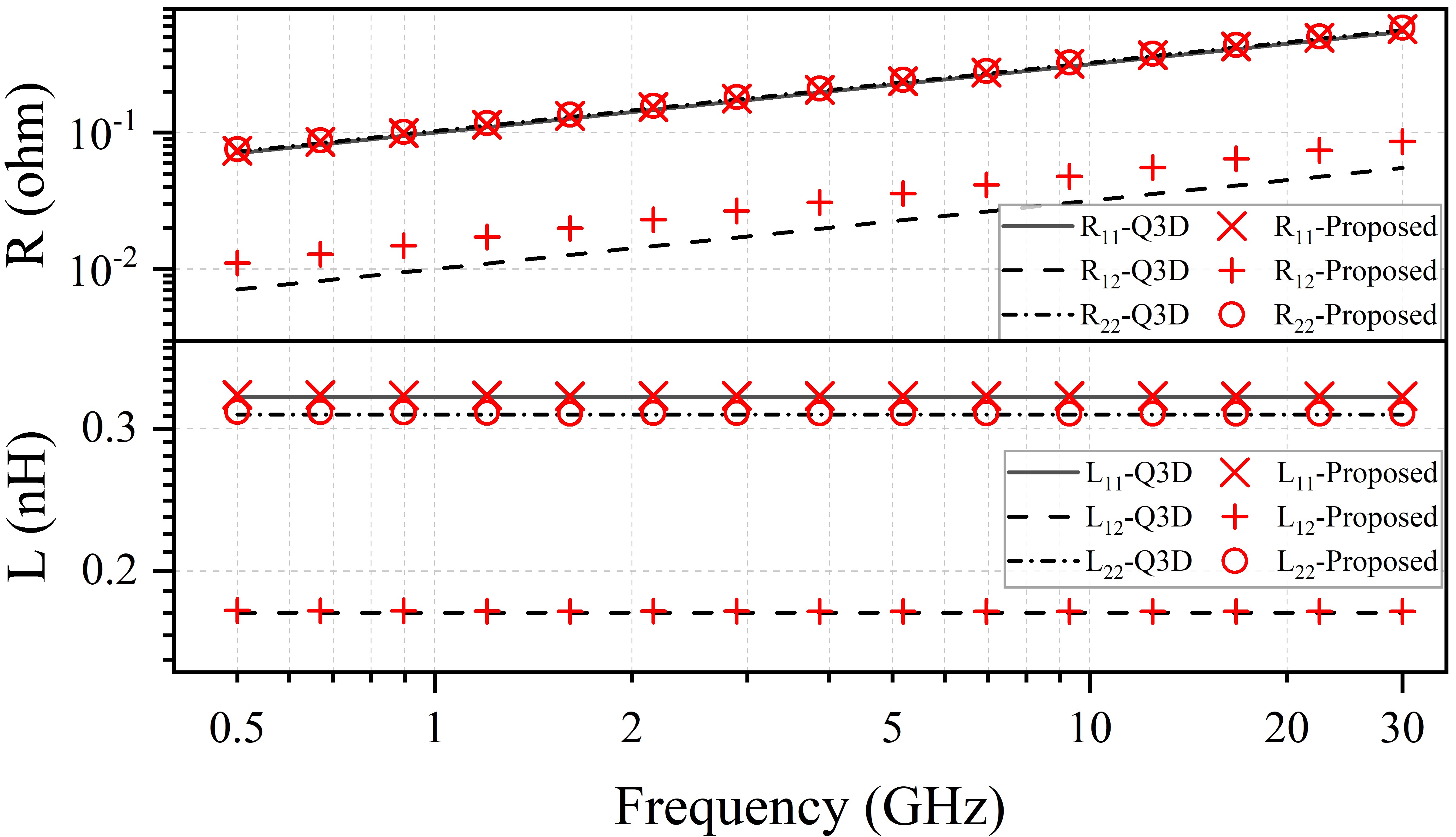}
	\caption{The self- and mutual- resistance and inductance obtained from the the Ansys Q3D and the proposed formulation.}
	\label{Case2_RL}
\end{figure}
\begin{figure}
	\centering
	\includegraphics[width=0.39\textwidth]{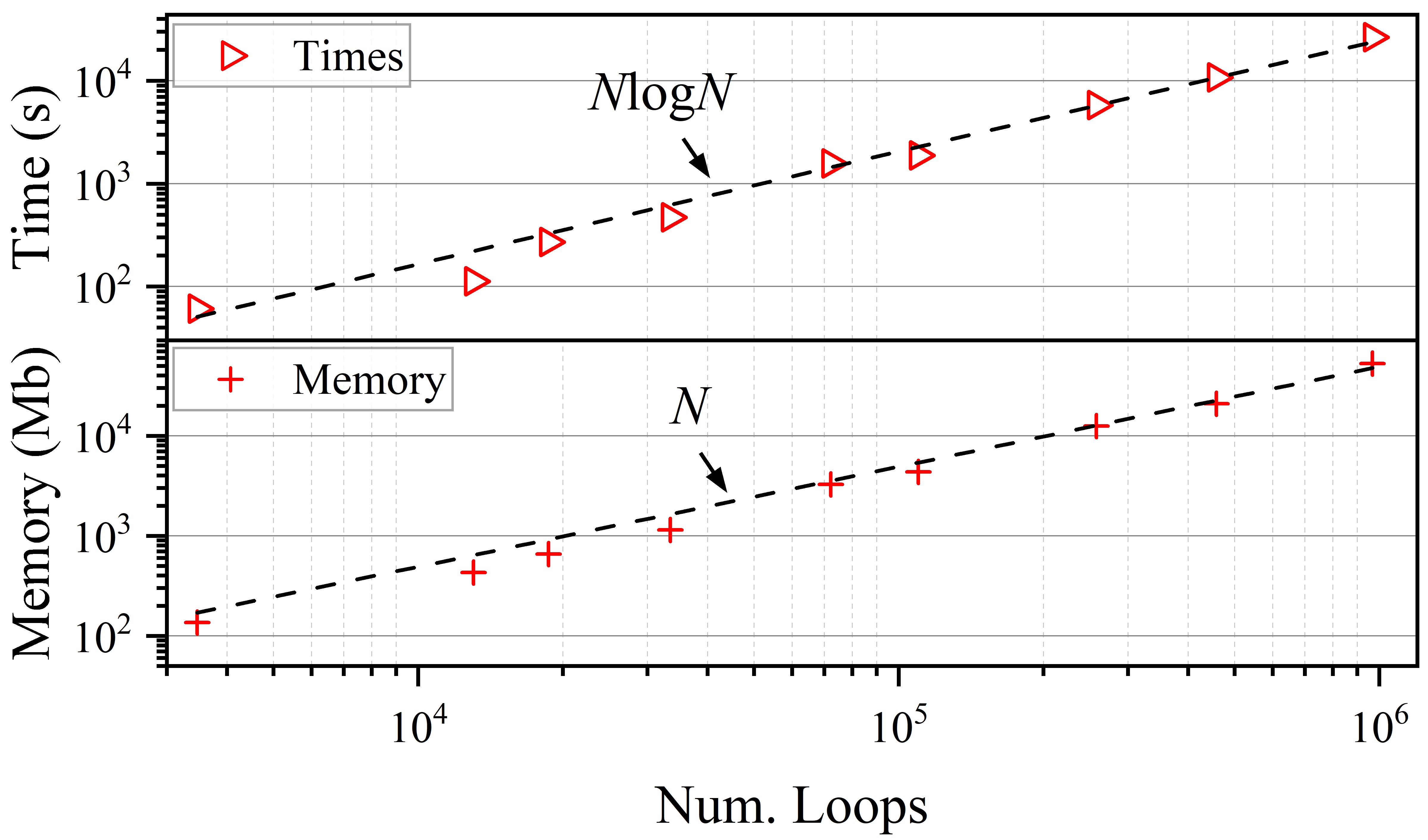}
	\caption{The runtime and memory consumption when different meshes are used to discretize the bounding wire.}
	\label{Case2_TM}
\end{figure}
\begin{figure}
	\centering
	\includegraphics[width=0.4\textwidth]{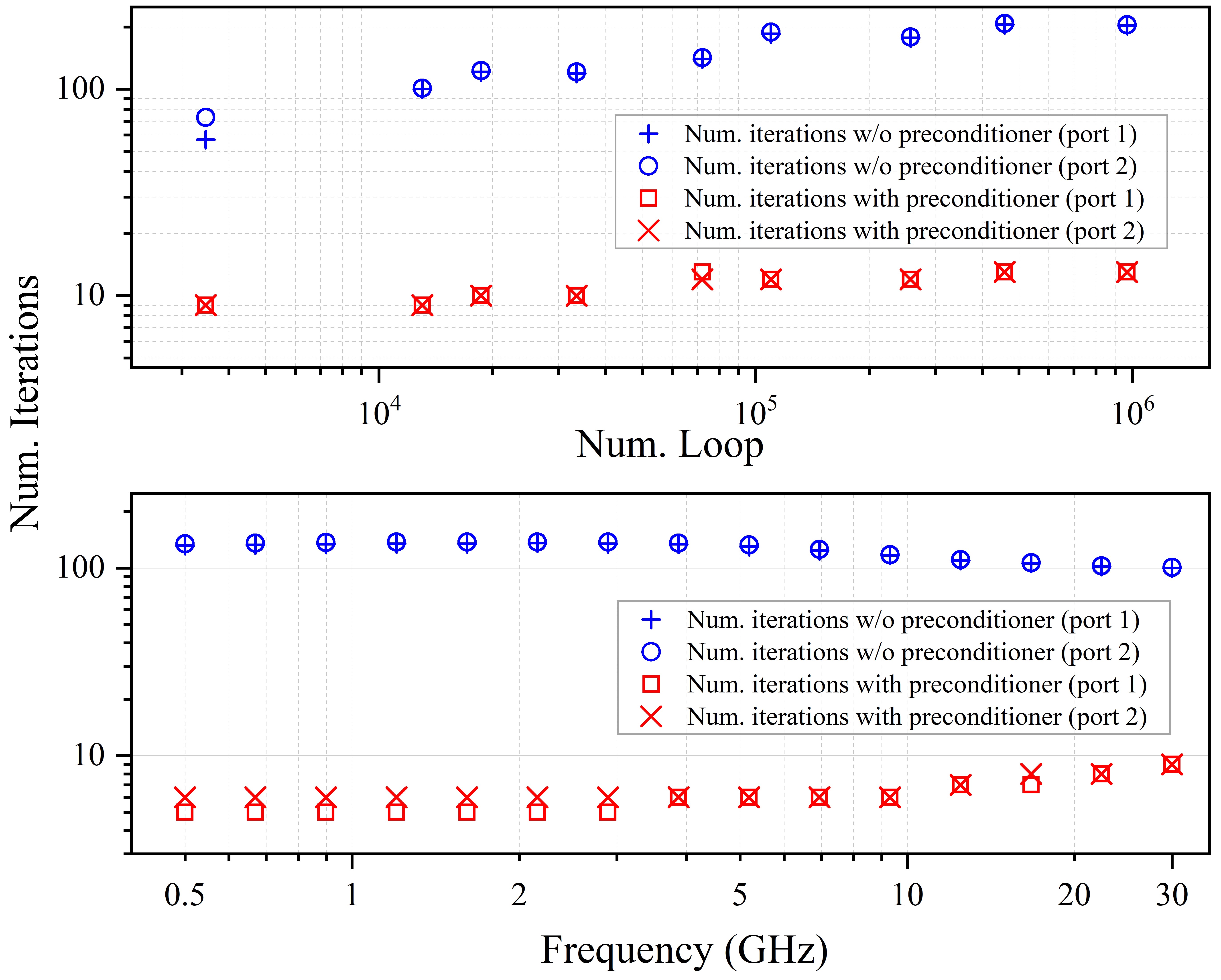}
	\caption{The number of iterations for GMRES with and without the preconditioner versus the frequency and meshes.}
	\label{Case2_It}
\end{figure}

The bounding wire array with 52 copper interconnects is considered here. The dimension of this structure is 1.2 mm $\times$ 1.2 mm $\times$ 0.15 mm, and the thickness of each interconnect is 10 um. The ports are defined on the ends of two adjacent bounding wires as shown in Fig. \ref{Case2_Model}. The frequency is set as 25 GHz, and different numbers of unknowns are used to solve this problem. To demonstrate the accuracy, we swept the frequency from 0.5 GHz $\sim$ 30 GHz, and calculated the resistance and inductance, in which 13,041 loops are found to construct the matrix equations. In Fig. \ref{Case2_RL}, results calculated by the proposed formulation and the Ansys Q3D are compared with each other. For both the self-inductance and mutual inductance, the two solvers show excellent agreement with each other for all sampling frequency, and the relative error is less than 2\%. The self-resistance also agrees well for the two solvers, but large difference occurs for the mutual resistance, which is about 25\%. In fact, the mutual resistance contributes little to the loop resistance due to their small values. The resource consumption including the runtime and memory is illustrated as shown in Fig. \ref{Case2_TM}. As the number of loops increases, the runtime and memory consumption also approximately match $O(NlogN)$ and $O(N)$.

To demonstrate the effectiveness of the preconditioner, the number of iterations as the number of loops and the frequency is recorded and illustrated in Fig. \ref{Case2_It}. As the number of loops increases, the number of iterations also increases with and without the preconditioner. However, the number of iterations increases very slowly when the preconditioner is used. For example, there is 9 iterations for 3,470 loops and 13 iterations for 967,909 loops. In general, the frequency has very little effect on the number of iterations. The overall number of iterations slightly decreases without the preconditioner and slightly increases with it when the frequency increases. In addition, as shown in Fig. \ref{Case2_Model}, the distribution of surface current and electric potential are illustrated when Port\#2 are excited by a voltage source. The scenario of the accumulation of current on the edges can be observed.
 
\subsection{Interconnects in A Real-life Circuit}
\begin{figure}
	\centering
	\includegraphics[width=0.49\textwidth]{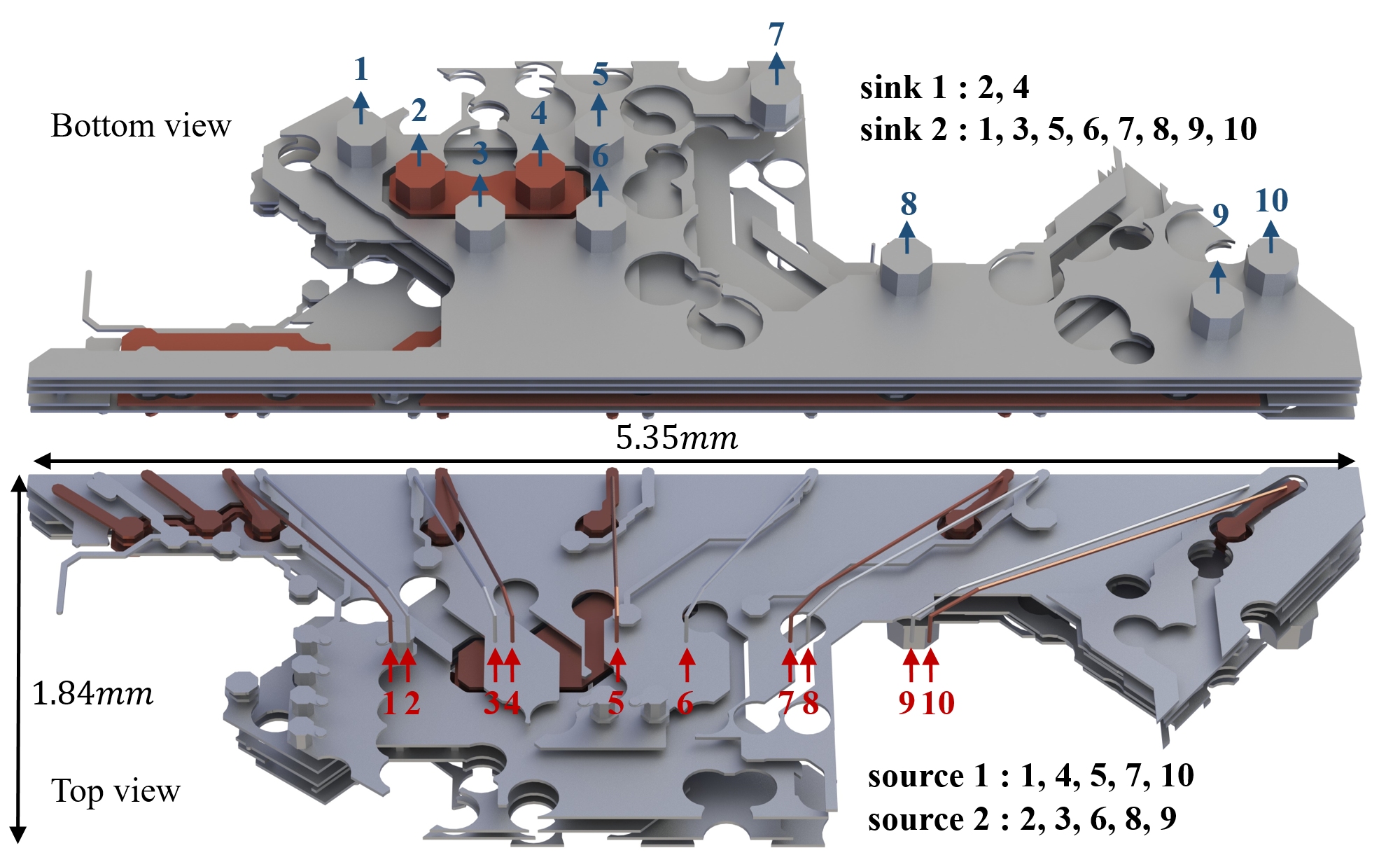}
	\caption{The bottom view and top view of a real-life circuit with several vias and the excitation configuration.}
	\label{Case3_Model}
\end{figure}
\begin{figure}
	\centering
	\includegraphics[width=0.4\textwidth]{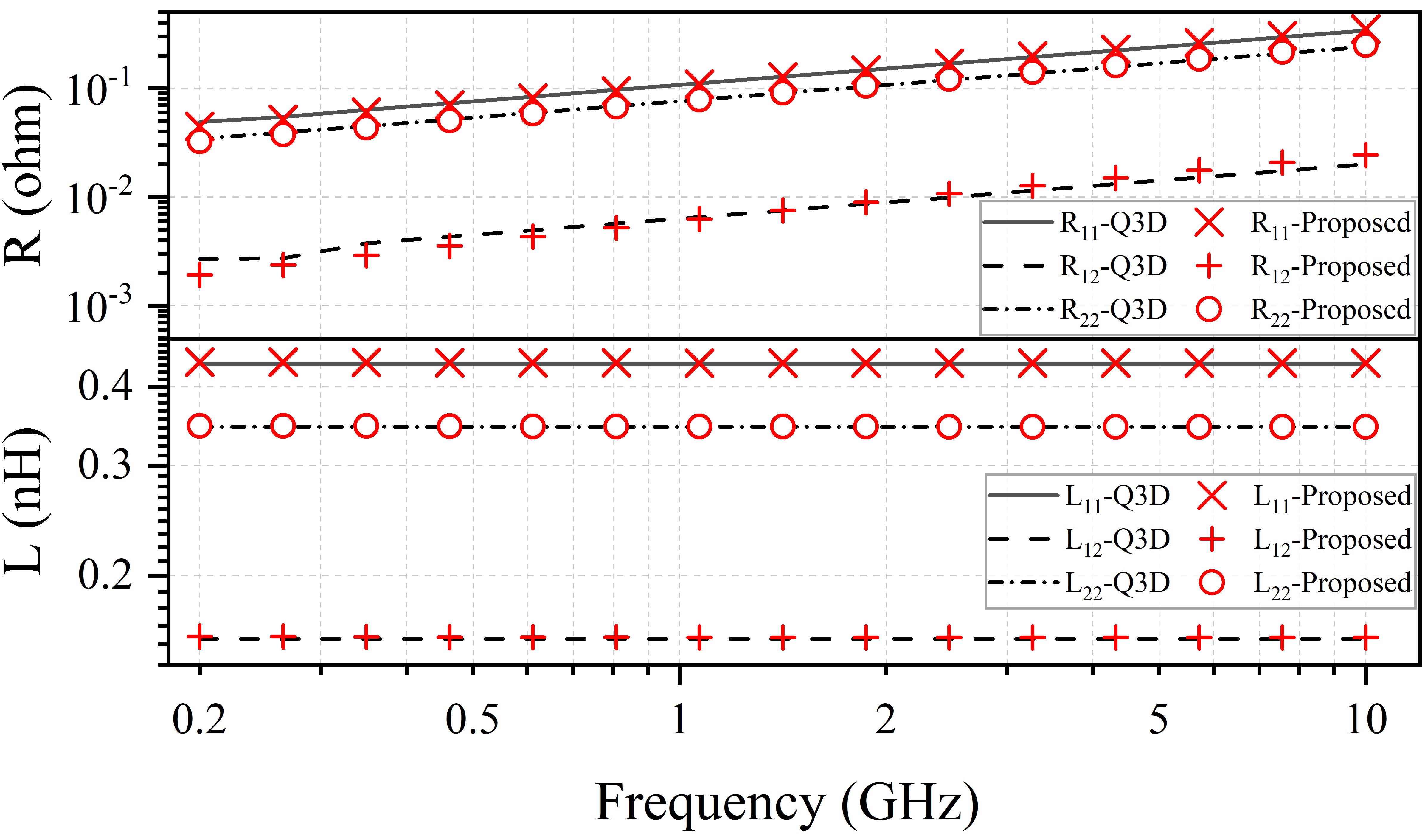}
	\caption{The self- and mutual- resistance and inductance obtained from the Ansys Q3D and the proposed formulation.}
	\label{Case3_RL}
\end{figure}
\begin{figure}
	\centering
	\includegraphics[width=0.39\textwidth]{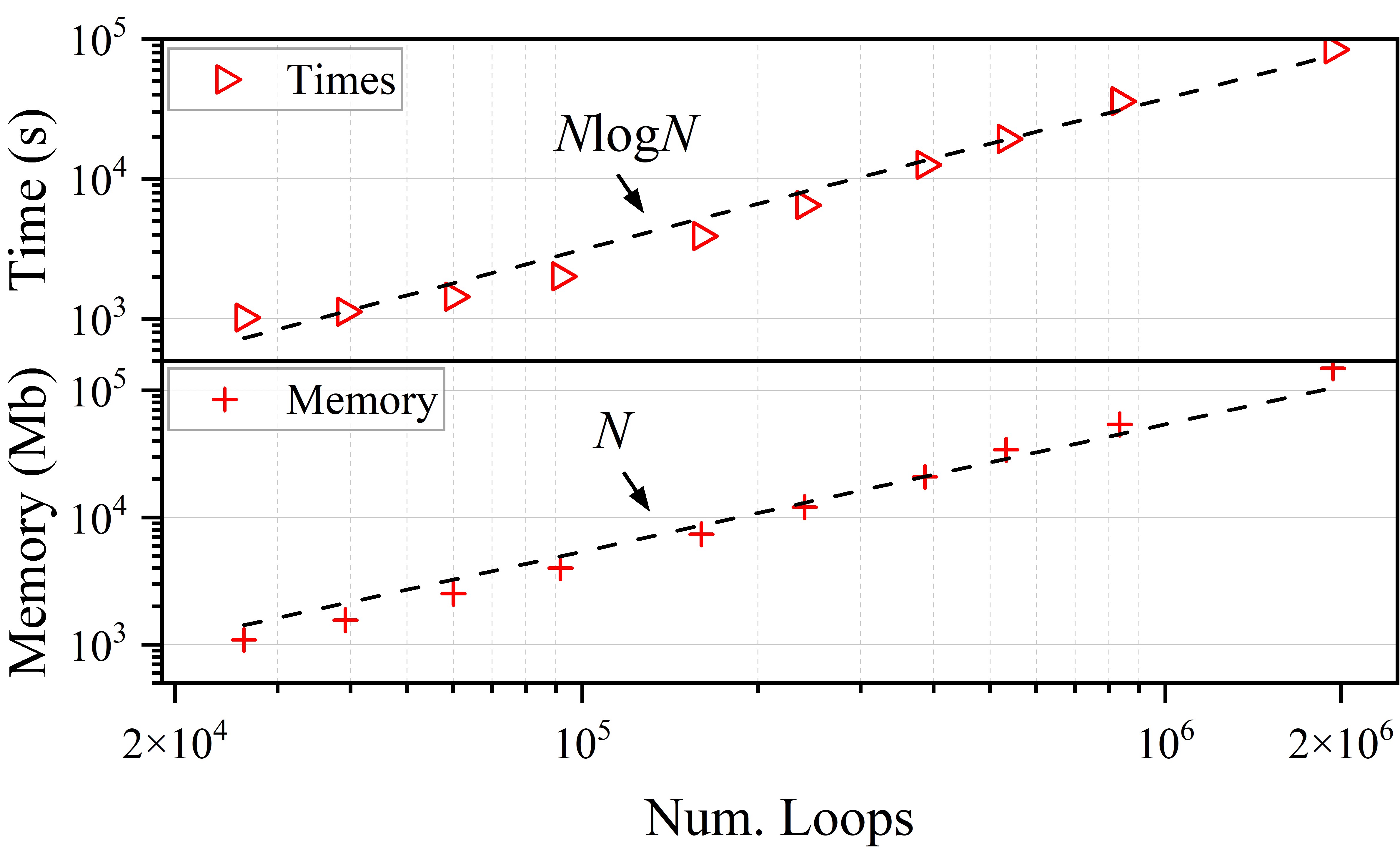}
	\caption{The time and memory consumption when different meshes are used to discretize the circuit.}
	\label{Case3_TM}
\end{figure}
\begin{figure}
	\centering
	\includegraphics[width=0.4\textwidth]{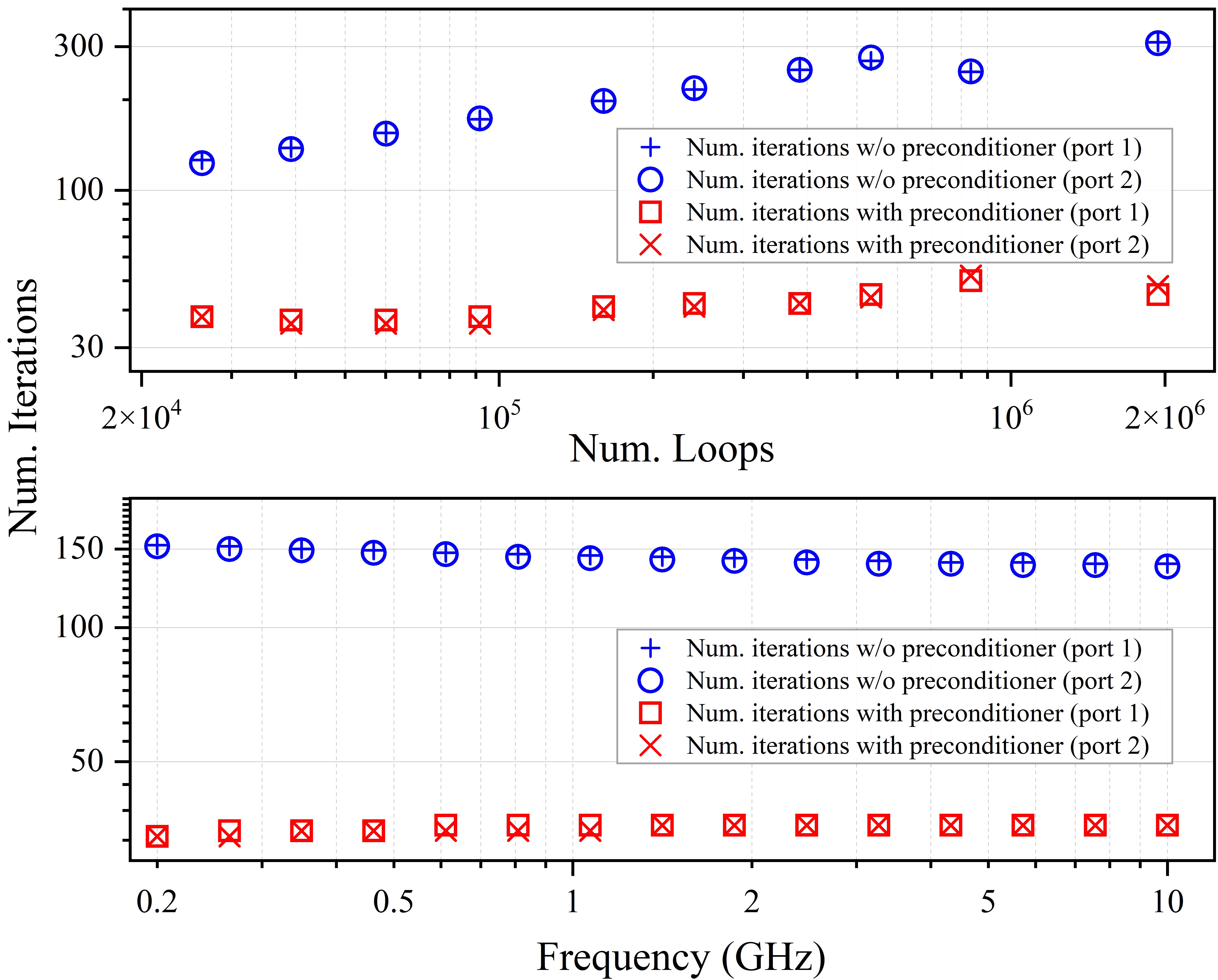}
	\caption{The number of iterations for the GMRES with and without the preconditioner versus the frequency and meshes.}
	\label{Case3_It}
\end{figure}
\begin{figure*}
	\centering
	\includegraphics[width=0.9\textwidth]{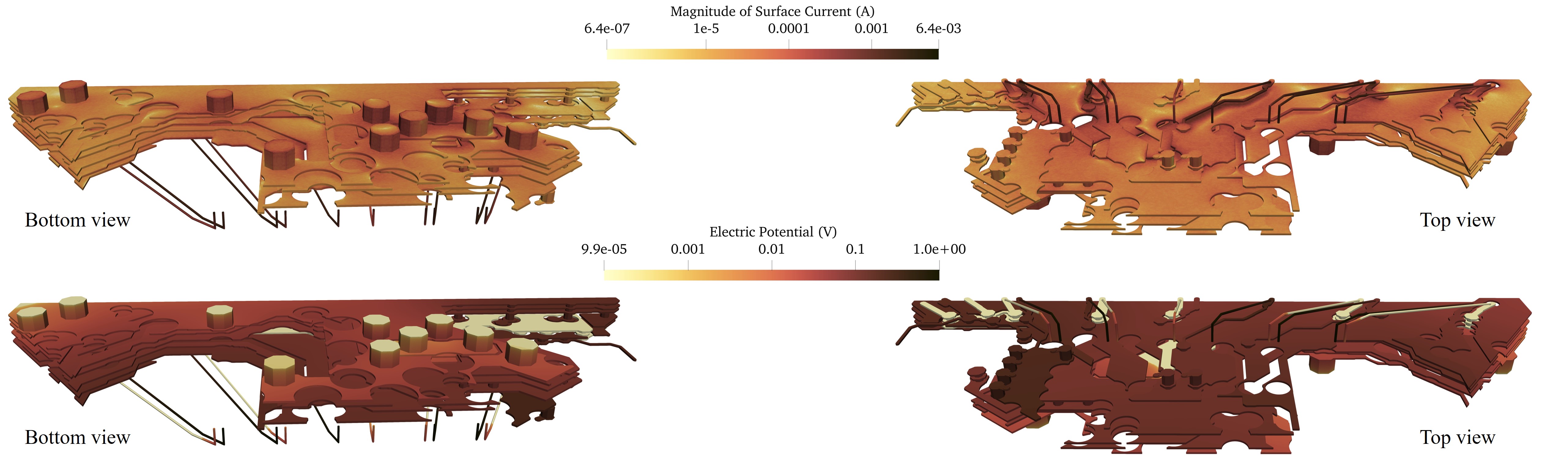}
	\caption{The magnitude of the surface current and the distribution of electric potential when a voltage source is applied on Port\#2.}
	\label{Case3_PC}
\end{figure*}

In this example, a real-life circuit with complex interconnects is considered and two ports are defined. The sources are defined on facets at the end of bounding wires extending from the circuit, and Source\#1 consists of $1^{\text{st}}$, $4^{\text{th}}$, $5^{\text{th}}$, $7^{\text{th}}$, $10^{\text{th}}$ facets while Source\#2 is composed by others. The sinks are set on the ubumps at the bottom of circuit as shown in Fig. \ref{Case3_Model}. We excited two ports in turn, and calculated the self- and mutual- resistance and inductance in the frequency range of 0.2 GHz $\sim$ 10 GHz. 77,352 triangular facets are used to discretize the surface, and 39,220 loops are required to construct the matrix equations. Fig. \ref{Case3_RL} shows results obtained from the proposed formulation and the Ansys Q3D. For the self- and mutual- resistance and inductance, results calculated by two solvers show excellent agreement with each other.

Fig. \ref{Case3_TM} shows the runtime and memory consumption when different meshes were used and the working frequency was fixed at 10 GHz. It can be found that the runtime approximately matches $O(NlogN)$, but it is slightly larger than $O(NlogN)$ as the unknowns increase. The main reason is that the runtime may match the complexity of $O(NlogN)$ for each iteration in the GMRES, but the number of iterations increases with the number of unknowns, which leads to a slightly larger runtime. As for the memory consumption, further optimization for our in-house solver is stilled required to be carried out to match $O(N)$ well for large-scale problems.

Similar to the previous two examples, the effectiveness of the preconditioner is investigated by the number of iterations, and we set the convergent tolerance as $10^{-3}$. The top subfigure of Fig. \ref{Case3_It} focuses on the number of iterations over meshes when the frequency is fixed at 10 GHz. The bottom subfigure relates the number of iterations to the frequency. There are 39,220 loops to construct the matrix. It can be found that the preconditioner significantly accelerates the convergence of the GMRES. Especially, the number of iterations increases very slowly when the number of unknowns increases if the preconditioner is applied. There are 38 iterations with preconditioner and 123 iterations without it for 26,256 unknowns. However, the scenario with the preconditioner requires 48 iterations for 1,938,754 unknowns while 310 iterations are required for the same unknowns without it. In addition, as shown in Fig. \ref{Case3_PC}, we calculated the magnitude of surface current and the distribution of electric potential when a voltage source is applied on Port\#2. It can be found that the current accumulates on the slender bounding wires. Therefore, the electric potential goes down very fast on them.
\subsection{Large-scale PDN Modeling}
\begin{figure}
	\centering
	\includegraphics[width=0.35\textwidth]{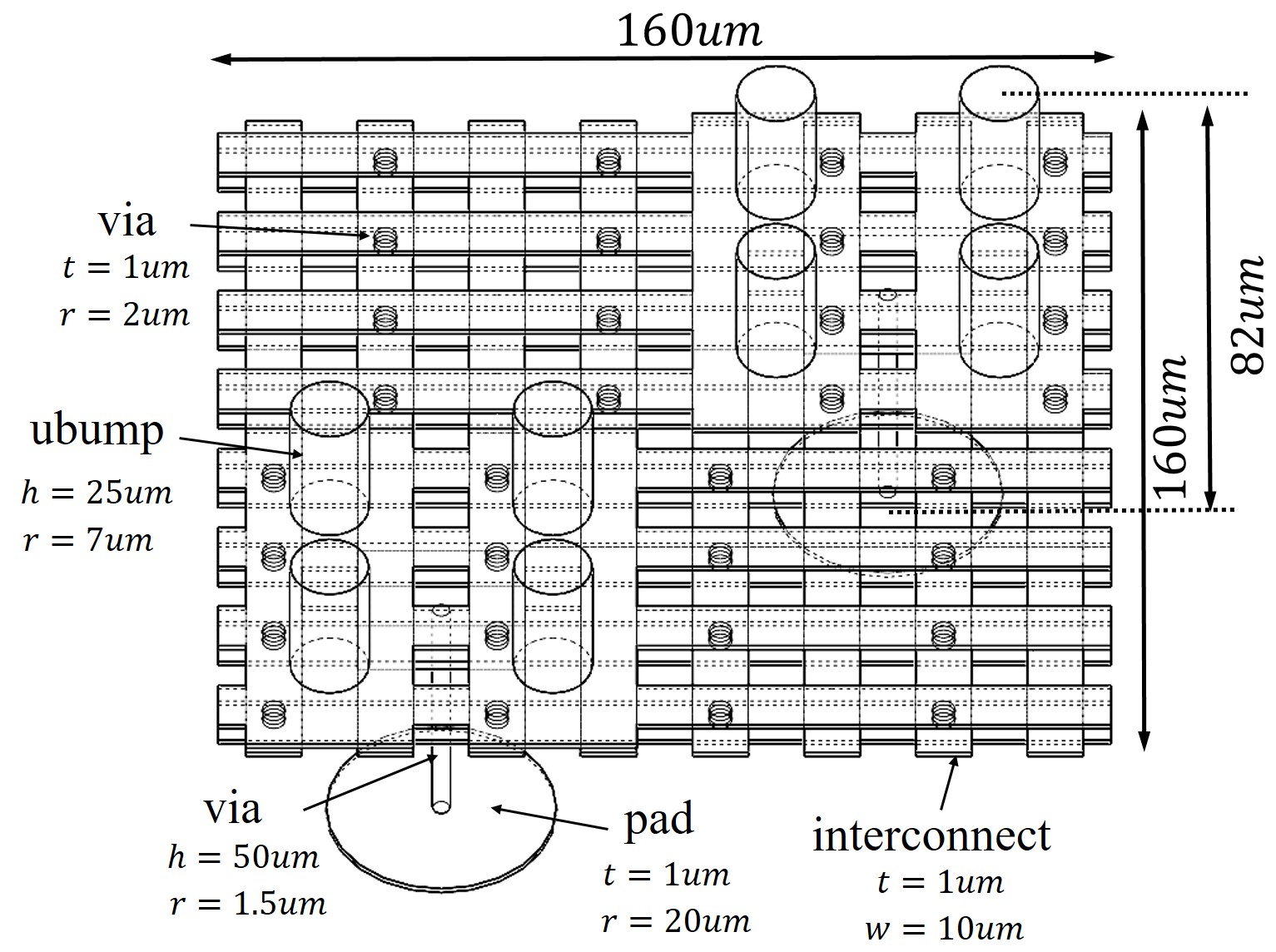}
	\caption{Geometrical details of the PDN with the size of 160 um $\times$ 160 um $\times$ 82 um.}
	\label{Case4_Model}
\end{figure}

\begin{figure}
	\begin{minipage}[h]{0.45\linewidth}
		\centerline{\includegraphics[scale=0.3]{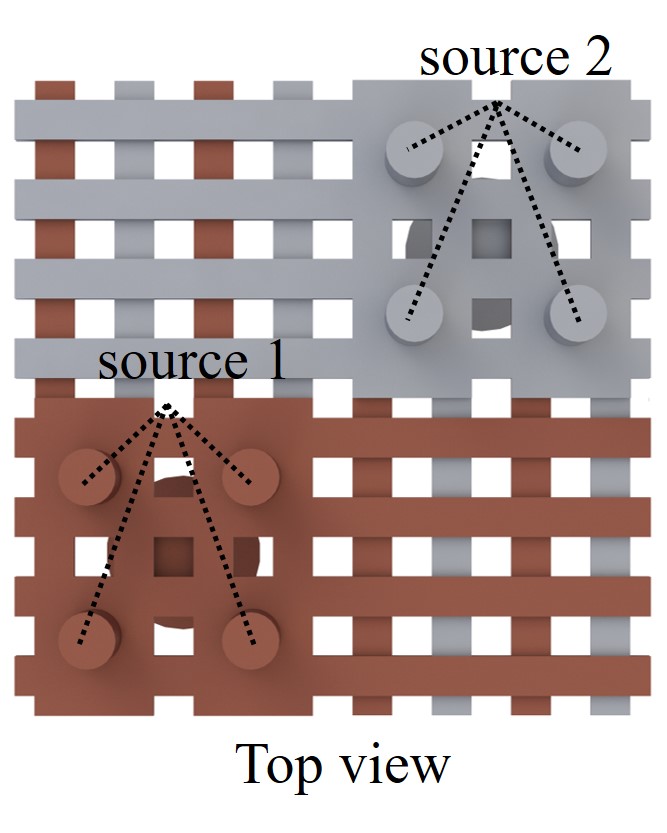}}
		\centerline{(a)}
	\end{minipage}
	\hfill
	\begin{minipage}[h]{0.45\linewidth}
		\centerline{\includegraphics[scale=0.3]{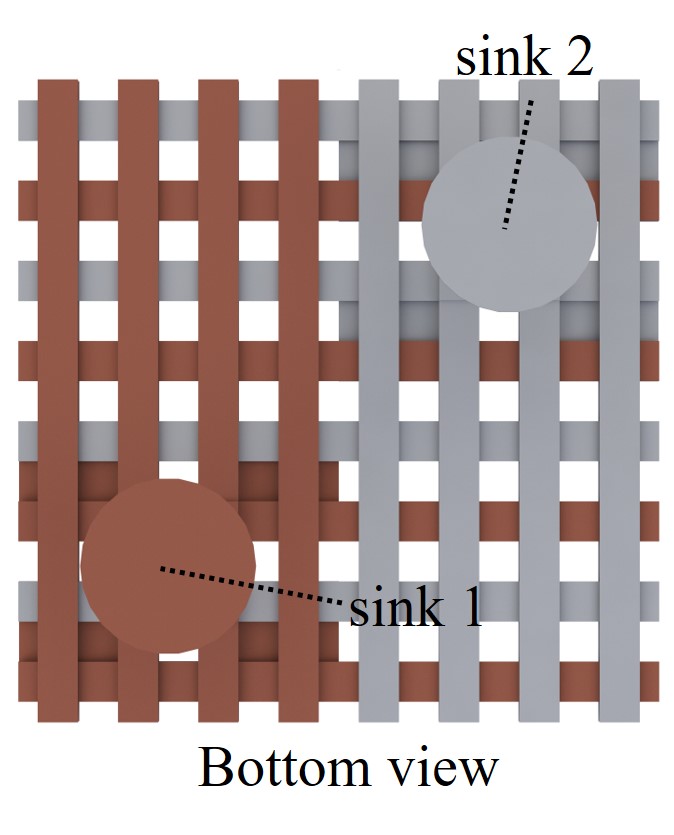}}
		\centerline{(b)}
	\end{minipage} 
	\caption{The top and bottom views of the PDN and the definition of ports.}
	\label{Case4_Port}
\end{figure}

In this example, the power distribution network (PDN) is considered to verify the scalability of the proposed formulation. As shown in Fig. 16, the cross rectangular interconnects with four ubumps on the top and one pad on the bottom are considered, and the structure has a dimension of 16 um$\times$16 um$\times$8.2 um. The interconnects of different layers are connected through a number of vias and the pad is coupled to the interconnects through a slender via. The structure in Fig. \ref{Case4_Model} is considered as a unit and we extend it to 2$\times$2, 4$\times$4, 6$\times$6 arrays. The sources are defined on the ubumps on the two contactless parts and the sinks are set on the bottom of pads as shown in Fig. \ref{Case4_Port}. The resistance and inductance were calculated at 30 GHz. The simulation configuration, results and resource consumption are presented in Table \ref{table2}.

\begin{figure*}
\begin{minipage}[h]{0.1\linewidth}
	\centerline{\includegraphics[scale=0.061]{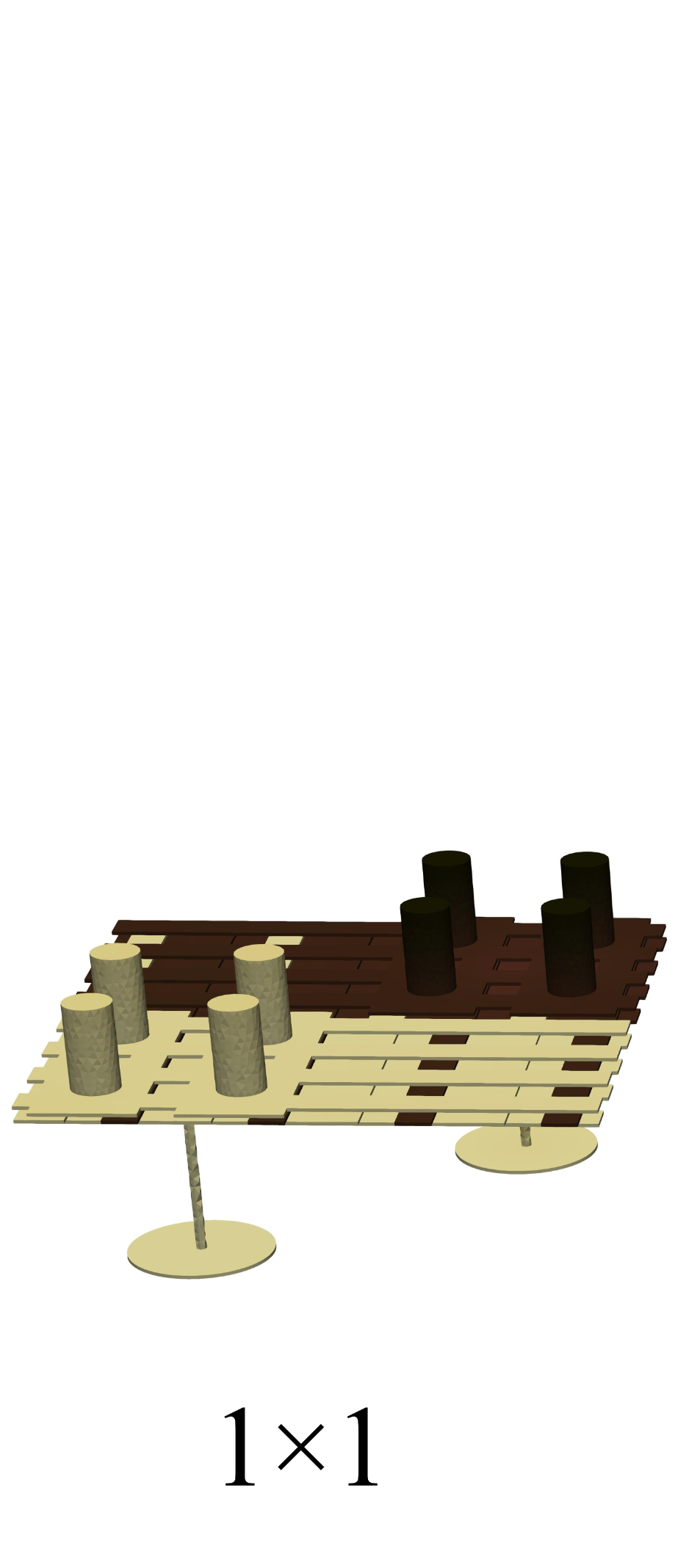}}
	\centerline{(a)}
\end{minipage}
\hfill
\begin{minipage}[h]{0.2\linewidth}
	\centerline{\includegraphics[scale=0.061]{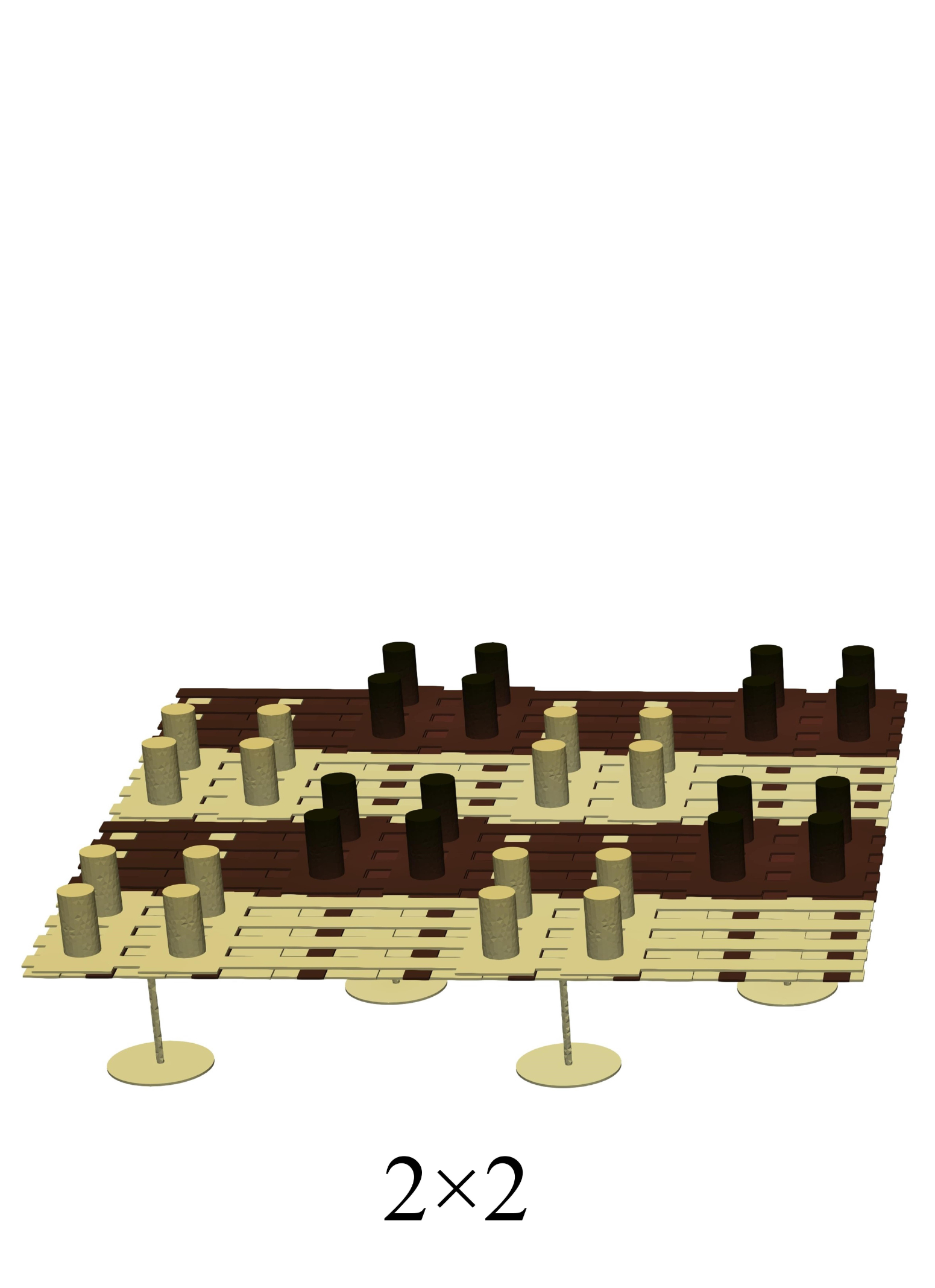}}
	\centerline{(b)}
\end{minipage}
\hfill
\begin{minipage}[h]{0.22\linewidth}
	\centerline{\includegraphics[scale=0.061]{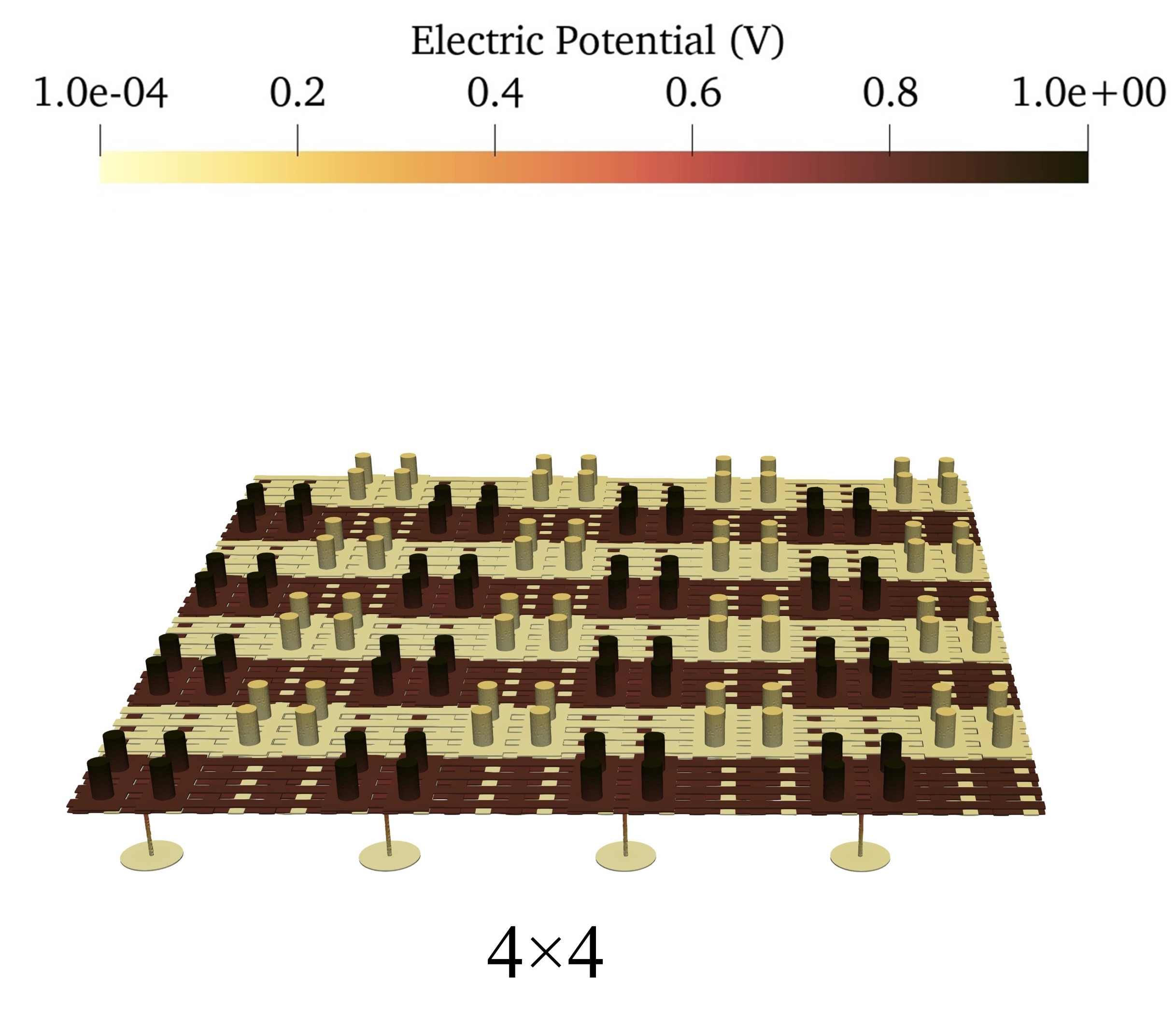}}
	\centerline{(c)}
\end{minipage} 
\hfill
\begin{minipage}[h]{0.4\linewidth}
	\centerline{\includegraphics[scale=0.074]{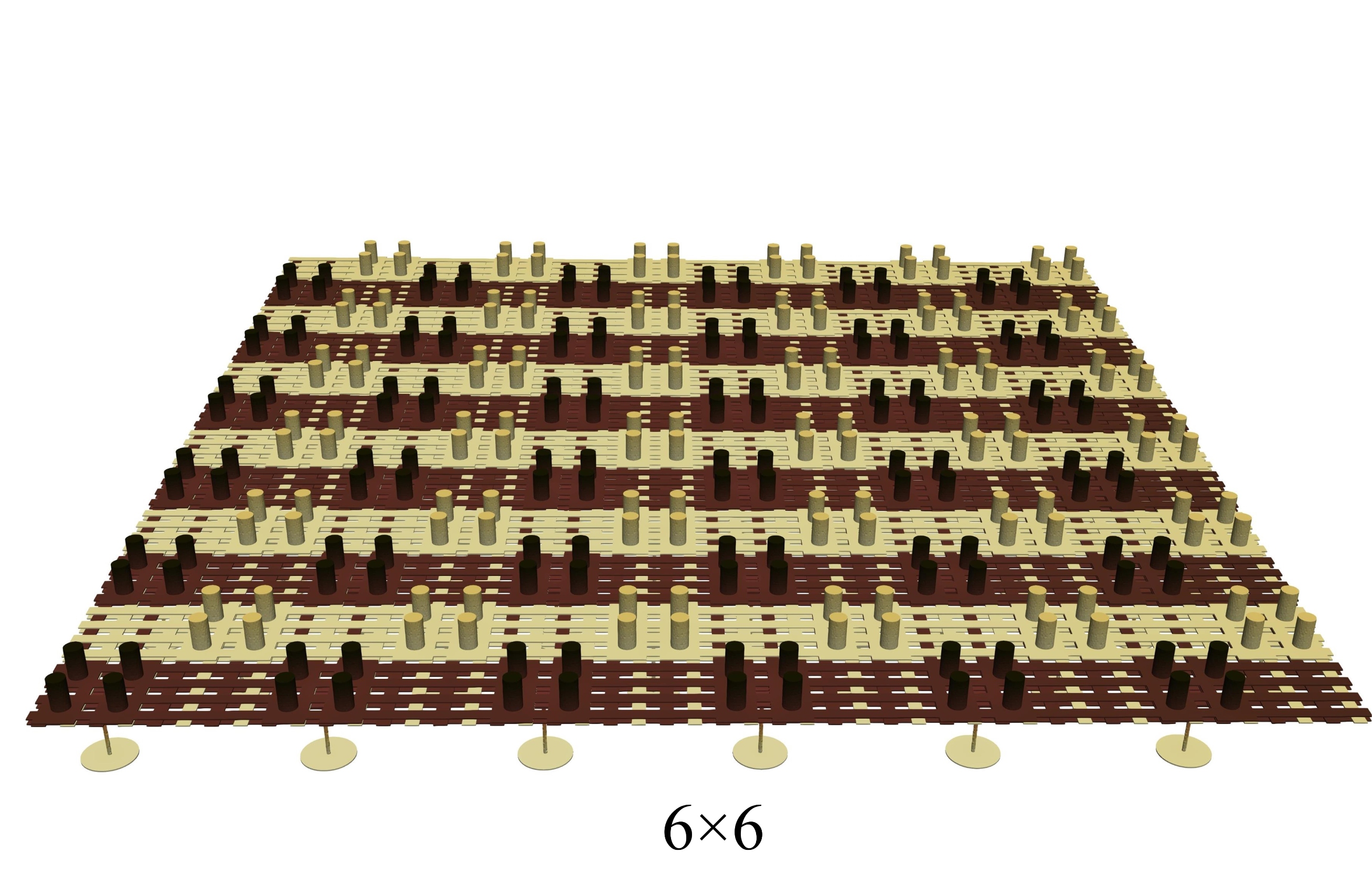}}
	\centerline{(d)}
\end{minipage} 
\caption{The Illustration of 1$\times$1, 2$\times$2, 4$\times$4 and 6$\times$6 PDNs. The electric potential is presented when Port\#1 is excited for 1$\times$1, 2$\times$2 PDNs and Port\#2 is excited for 4$\times$4, 6$\times$6 PDNs.}
\label{PDN66}
\end{figure*}

The first two rows record the overall number of triangles and loops used to construct the PDN. Next the runtime, memory, number of iterations with and without the preconditioner are presented when the convergent tolerance is set to $10^{-3}$, in which we summed the number of iterations for two solutions to the Port\#1 and Port\#2. The preconditioner still works very well for this example, and it significantly reduces the number of iterations by one order. In addition, the loop resistance and inductance are calculated by the proposed formulation and the Ansys Q3D, which are defined as $R_o = R_{11} – 2R_{12} + R_{22}$ and $L_o = L_{11} – 2L_{12} + L_{22}$. For the loop inductance, the proposed formulation shows excellent agreement with that from the Ansys Q3D, and the relative error is less than 4\% for the four cases. The loop resistance has a slightly larger error, still only 7\%. Fig. \ref{PDN66} gives the illustration of the PDNs with different dimensions and the distribution of electric potential is shown, in which the 1$\times$1, 2$\times$2 arrays are excited on Port\#1 and 4$\times$4, 6$\times$6 arrays are excited on Port \#2. There is little change in electric potential for the upper part of PDNs, and the electric potential goes down very fast on the vias connected to the pads.
\renewcommand\arraystretch{1.2}
\begin{table}	
	\begin{center}		
		\caption{Simulation configuration, resource consumption, loop resistance and inductance obtained from the proposed formulation and Ansys Q3D}\label{table2}
		\begin{tabular}{|l|c|c|c|c|}
			\hline
			\multicolumn{5}{|c|}{$f=30$ GHz, tol = $10^{-3}$}\\
			\hline
			\hline
			\,            & 1$\times$1          & 2$\times$2                  & 4$\times$4          & 6$\times$6 \\
			\cline{1-5}
			\rule{-3pt}{9pt}
			Num. triangles     & 62,590              & 261,928                        & 1,051,950           & 1,933,368 \\
			\cline{1-5}
			\rule{-3pt}{9pt}
			Num. Loops         & 33,047              & 138,735                      & 558,084             & 1,025,824\\
			\cline{1-5}
			\rule{-3pt}{15pt}
			\makecell[l]{Num. iterations  \\ with preconditioner} & 21              & 31                          & 44             & 54\\
			\cline{1-5}
			\rule{-3pt}{15pt}	
			\makecell[l]{Num. iterations  \\ w$\backslash$o preconditioner} & 230              & 423                          & 804             & 1,811\\
			\cline{1-5}
			\rule{-3pt}{9pt}
			\makecell[l]{Runtime (s)} & 485              & 3,566                        & 18,401             & 47,220\\
			\cline{1-5}
			\rule{-3pt}{9pt}
			\makecell[l]{Memory (Mb)} & 1,307              & 9,948                          & 104,186            & 222,494\\
			\hline
			\hline
			\rule{-3pt}{9pt}
			\makecell[l]{$R_o$-proposed (ohm)} & 0.68              & 0.17                          & 0.043             & 0.020\\
			\cline{1-5}
			\rule{-3pt}{9pt}
			\makecell[l]{$L_o$-proposed (nH)} & 0.083              & 0.020                         & 0.0050             & 0.0022\\
			\cline{1-5}
			\rule{-3pt}{9pt}
			\makecell[l]{$R_o$-Ansys Q3D (ohm)} & 0.70              & 0.18                      & 0.045            & 0.021\\
			\cline{1-5}
			\rule{-3pt}{9pt}
			\makecell[l]{$L_o$-Ansys Q3D (nH)} & 0.083              & 0.021                          & 0.0052             & 0.0022\\
			\cline{1-5}
		\end{tabular}
	\end{center}
\end{table}


\section{Conclusion}
In this paper, an MQS-SIE formulation with the loop analysis is proposed to model interconnects in packages at high frequencies. The triangle discretization are used to support the surface current flowing in any direction. The graph theory in circuit analysis is introduced to construct the independent and complete loop equations. By transferring the branch quantities to loop quantities, the dimension of the system matrix can reduce by about 80\%. In addition, the pFFT algorithm is successfully carried out for the proposed formulation, and an efficient preconditioner is developed to speed up the convergence in the GMRES. The numerical examples verify the scalability and effectiveness of the proposed formulation, accurate results can be obtained compared with those from the industrial solver Ansys Q3D. 

However, the uniform grid option required by the traditional pFFT algorithm will cause the waste of computational resource, especially for the multiscale structures. Therefore, the optimization of the pFFT algorithm, like the hierarchical algorithm, will be our future work.

\end{document}